\documentclass{aa}  
\usepackage[colorlinks=true,     linkcolor=blue, citecolor=blue, filecolor=blue, urlcolor=blue]{hyperref}
\usepackage{graphicx}
\usepackage{txfonts}
\begin{document}

   \title{Intermediate- and high-resolution optical spectroscopy on the nascent planetary nebula Vy\,2-2: Evidence of binary-star evolution}

 \subtitle{}

   \author{Rold{\'a}n A. Cala \inst{1},
          Luis F. Miranda\inst{1}, %\fnmsep\thanks{Just to show the usage of the elements in the author field},
      	   Jos{\'e} F. G{\'o}mez\inst{1},
         Lucero Uscanga \inst{2},
         \and
         J. Ricardo Rizzo\inst{3}}

   \institute{Instituto de Astrof{\'i}sica de Andaluc{\'i}a (IAA), CSIC, Glorieta de la Astronom{\'i}a s/n, 18008 Granada, Spain\\
              \email{roldancalab@gmail.com}
         \and
         Departamento de Astronomía, Universidad de Guanajuato, A.P. 144, 36000 Guanajuato, Gto., Mexico
         \and
         ISDEFE, Beatriz de Bobadilla 3, 28040 Madrid, Spain
%             University of Alexandria, Department of Geography, ...\\
%             \email{c.ptolemy@hipparch.uheaven.space}
%             \thanks{The university of heaven temporarily does not
%                     accept e-mails}
             }

   \date{Received ; accepted }

% \abstract{}{}{}{}{} 
% 5 {} token are mandatory
  \abstract{Planetary nebulae (PNe) tend to show spectacular non-spherical shapes whose origin is not fully understood. Crucial information about PNe formation can be obtained by studying their very first stages. We present and analyze new intermediate- and échelle high-resolution spectra of the nascent PN Vy\,2-2 to obtain its chemical abundances and a new value for its controversial abundance discrepancy factor (\textit{adf}), investigate the internal kinematics and ionization of the nebula, and infer its possible formation processes. Our intermediate resolution spectra reveal many previously unreported nebular recombination (e.g., N$^+$) and forbidden (e.g., [F\,{\sc iv}]) emission lines as well as stellar emission lines. Some of the newly detected nebular lines are important to determine the \textit{adf}. The derived abundances are subsolar and agree with previous values, while the new value of the \textit{adf}(O$^{2+}$)=13.6$\pm$2.6 indicates that Vy\,2-2 belongs to the group of PNe with unusually high \textit{adf}. Our morphokinematical and ionization structure analysis reveals that Vy\,2-2 consists of a relatively high-excitation central region and two low-excitation bipolar lobes. Our data does not show significant shock excitation in the main ionized shell of this PN. A distance of $\simeq$1.68 kpc to Vy\,2-2 is obtained from the published angular expansion of the central region and its expansion velocity in H$^+$. The inferred kinematical age of $\simeq$210\,yr supports that Vy\,2-2 is a nascent PN. The stellar emission lines suggest a \textit{wels}-like spectrum although more likely associated with an irradiated close companion than with the central star. Actually, the unusually high \textit{adf}, the observed nebular structures, presence of fluorine, among others, point to Vy2-2 hosting a post-common-envelope binary system. Chemical abundances and circumstellar mass estimates suggest an initial mass $\lesssim0.9\,M_{\odot}$ for the star progenitor of Vy\,2-2.}
%  \abstract
  % context heading (optional)
  % {} leave it empty if necessary  
 % {}
  % methods heading (mandatory)
  % {}
  % results heading (mandatory)
  % { }
  % conclusions heading (optional), leave it empty if necessary 
  % {}

   
\authorrunning{Cala, R.~A., et al.}
\titlerunning{the nascent planetary nebula Vy\,2-2} 
 
   \maketitle
%
%-------------------------------------------------------------------

\section{Introduction}

The evolution of stars with initial masses (M$_{\rm i}$) between $\simeq$0.8 and 8\,M$_{\odot}$ (e.g., the Sun) from the asymptotic giant branch (AGB) into the planetary nebula (PN) phase is one of the fastest and most astonishing processes in stellar evolution, resulting in the transformation of roughly spherical envelope into a plethora of non-spherical morphologies \citep[][]{man96}. The identification of nascent PNe is thus crucial to investigate the first stages of the PN formation. In particular, OH maser-emitting PNe (OHPNe) are considered to be nascent PNe \citep[][]{usc12, cal25} because OH masers tend to disappear in timescales of $\sim$10$^{3}$\,yr after the AGB \citep{lew89, uno21}, while, in comparison, the PN phase starts 10$^{3}$-10$^{4}$ yr after the AGB, depending on M$_{\rm i}$ \citep[]{blo95, mil16}. To date, only seven OHPNe have been identified: Vy 2-2, IRAS\,17393--2727, K\,3--35, IRAS\,17347--3139, JaSt\,23, IRAS\,16333--4807, and Vo\,1 \citep[][]{sea83, mir01, usc12, qia16a, cal25}. 

% ionized by central stars of different spectral types \citep[][]{wei20}

Vy\,2-2 \citep[PN G045$-$02.7;][]{vys45} is the first PN in which maser emission was detected. \cite{dav79} reported the detection of the OH maser line at 1612\,MHz towards this PN, based on Arecibo observations. \cite{sea83} confirmed the association of the radio continuum and 1612\,MHz emission from Vy\,2-2 using accurate interferometric positions. Afterwards, \cite{gom16} reported the presence of shock-excited OH maser transition at 1720\,MHz. Water and silicon monoxide masers have not been detected to date in this PN \citep[][]{gom90, deG04, cal24}.

Vy 2-2 is also peculiar because the ionic abundances obtained from optical recombination lines (ORLs) are higher than those obtained from collisionally excited lines (CELs) by an unusually large factor. The difference of ionic abundances obtained with these two types of lines is usually quantify with the abundance discrepancy factor (\textit{adf}), defined as the O$^{2+}_{\rm ORL}$/O$^{2+}_{\rm CEL}$ abundance ratio, and has a typical value of \textit{adf}(O$^{+2}$)$\lessapprox$5 among PNe \citep{wes05, wes18}. For Vy\,2-2, an initial value of \textit{adf}(O$^{+2}$)$\simeq$11.8 has been reported by \cite{wes05}. Such a high \textit{adf} has been attributed to the inclusion of H-deficient material \citep[][]{liu00, wes03}. Some authors have speculated that this material may be ejected during the interaction between the central star and a companion as found in PNe hosting a binary central star \citep{wes18, jac20}. However, a new value of \textit{adf}(O$^{+2}$)$\simeq$4.3 has been reported for this OHPN by \cite{reb22}, which is high but not unusual for PNe. It is obvious that a new independent determination of the \textit{adf} in Vy\,2-2 is required to solve this discrepancy and to properly interpret the object.

\begin{figure*}
	\centering
	\includegraphics[scale=0.405]{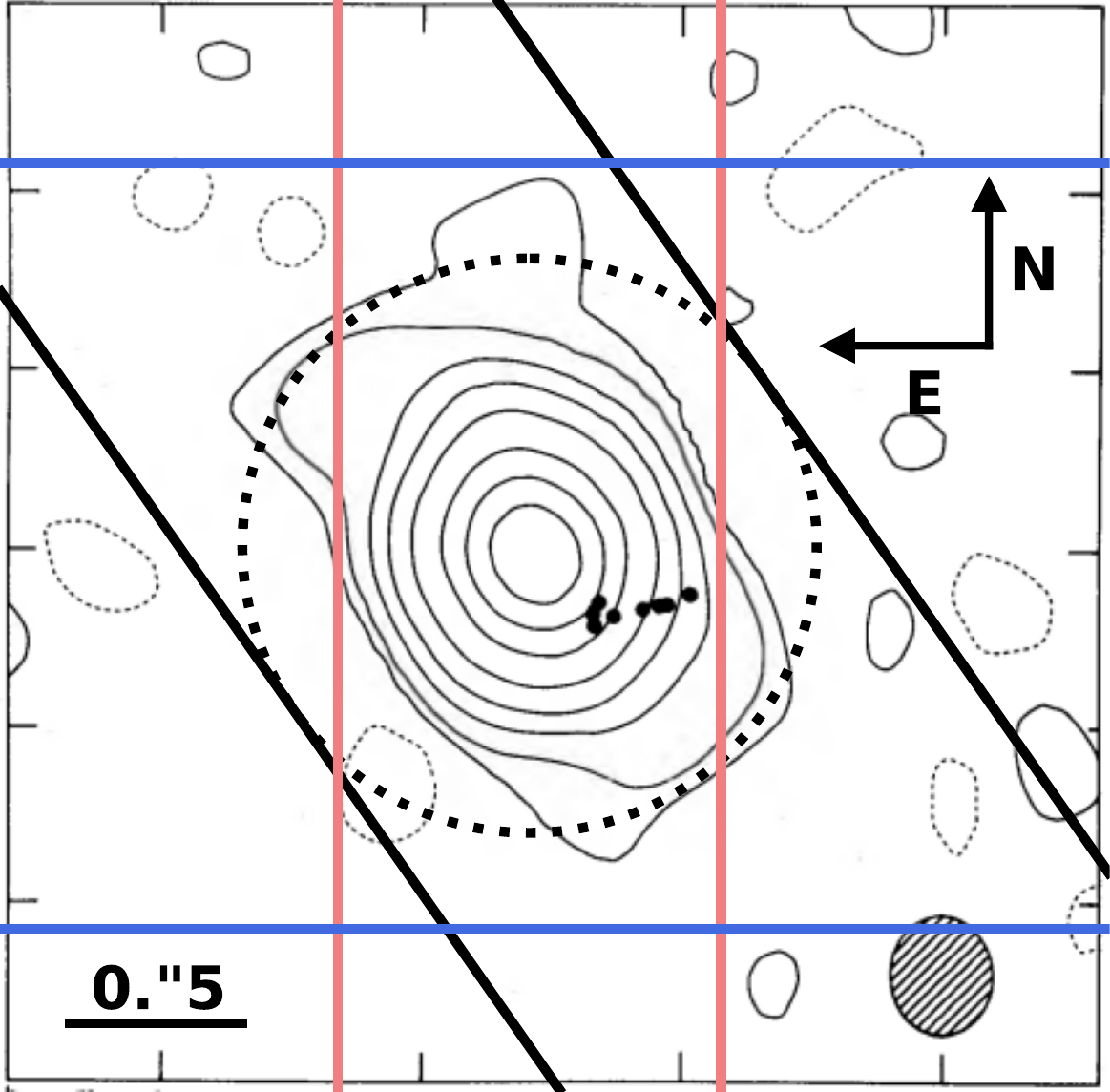}
    \hspace{0.5cm}
	\includegraphics[scale=0.4]{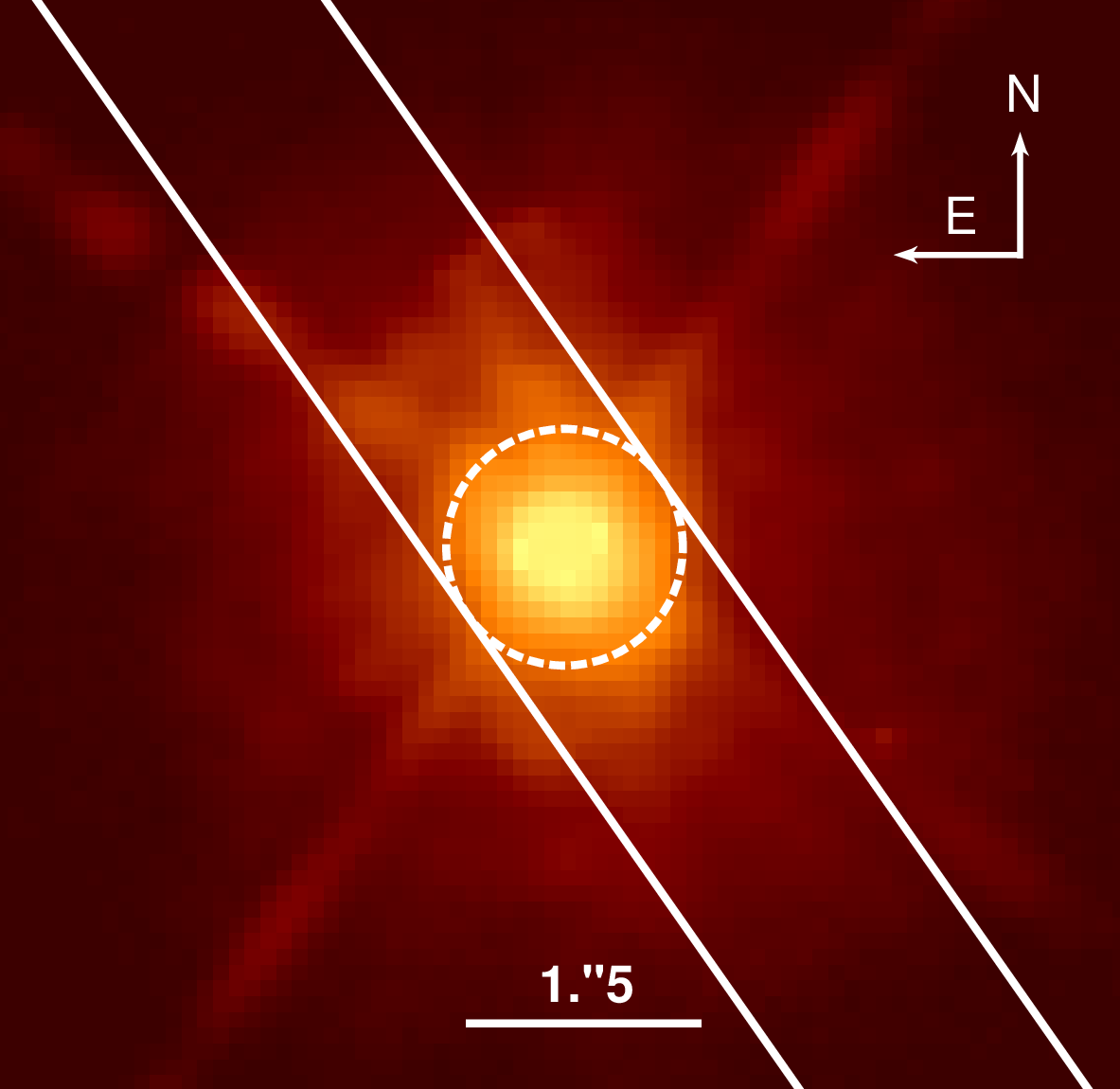}
	\caption{\textit{Left.} Contour map of Vy\,2-2 at 4.9\,GHz adapted from \cite{sea83}. The hatched ellipse at the bottom right represents the beamsize of the VLA interferometer. The black dots mark the positions of the 1612 MHz OH masers. The black solid lines and  dashed circle represent the long-slit of CAFOS (width = 1$\farcs$5) and the CARMENES fiber FOV (diameter = 1$\farcs$5), respectively. The red and blue lines represent the slits used by \cite{wes05} and \cite{reb22}, respectively. \textit{Right}. H$\alpha$ image of Vy\,2-2 obtained with the $HST$. The white solid lines and dashed circle correspond to the black lines and dashed circle in the left panel.}
	\label{fig:HST_caha}
\end{figure*}

The radio continuum emission of Vy\,2-2 is consistent with thermal (free-free) emission from photoionized gas. The flux density ($S_{\nu}$) dependence at radio frequencies $\nu$$<$14 GHz has a spectral index ($\alpha$, where $S_{\nu} \propto \nu^{\alpha}$) $\alpha$=1.30$\pm$0.07, which indicates a partially optically thick regime for the free-free emission \citep{sea83, chr98}. Figure\,\ref{fig:HST_caha} (left) shows the radio continuum map of Vy\,2-2 at 4.9 GHz adapted from \cite{sea83} (see their Fig. 1). It shows that this OHPN is composed of a very bright and compact central core surrounded by a weak bipolar formation with the major axis at position angle (pa) $\simeq$34$^{\rm o}$ \citep{sea83}. At frequencies $\nu$$>$14 GHz the radio continuum spectrum of Vy\,2-2 flattens to a value of $\alpha$$\simeq$$-$0.1, which is consistent with optically thin emission \citep{sea83}. Furthermore, the very bright core is resolved into a circular shell with a radius of $\simeq$0$\farcs$24 at 15 and 22\,GHz \citep{chr98}.  

Figure\,\ref{fig:HST_caha} (right) also presents the H$\alpha$ image of Vy\,2-2 obtained from the legacy archive of the Hubble Space Telescope (\textit{HST}). As in the radio images, it clearly shows a bright central core, and faint emission extending up to $\simeq$2$\farcs$3 from the center tracing a circular or slightly elliptical shell. \cite{mir91} (hereafter MS91) identified a bright, slowly expanding compact central core and a bipolar formation in Vy\,2-2 with the main axis at position angle PA$\simeq$34$^{\circ}$ by means of spatially resolved high-resolution optical long-slit spectroscopy. These structures probably correspond to those observed at 4.9\,GHz (Fig.\,1, left). The analysis by MS91 was based on the [N\,{\sc ii}]$\lambda$6583 emission line only, but it would be interesting to extend it to other emission lines to obtain a more complete picture of the internal kinematics and ionization structure of the object. Finally, \cite{ram16} found in the $HST$ images at least four concentric rings spaced by $\sim$0$\farcs$55 around the bipolar nebula. Moreover, \cite{kap26} recently found a pair of high-velocity jets or bullets along the North-South direction, indicating that  collimated outflows are also present in Vy\,2-2.

Other observations indicate the presence of carbon monoxide \citep[CO; ][]{hug96}, molecular hydrogen \citep[H$_{\rm 2}$; ][]{hora99}, a photodissociation region \cite[PDR; ][]{liu01}, silicate dust and H$_{2}$O ice \citep{mol02}, as well as the presence of fresh dredge up material \citep{ste15}. Silicon carbide (SiC) and polycyclic aromatic hydrocarbons (PAHs) have not been detected \citep[][]{rin02, coh05}. Therefore, the circumstellar environment of Vy\,2-2 is O-rich.

In this paper we present and analyze new intermediate- and high-resolution spectroscopic observations of the OHPN Vy\,2-2. The intermediate-resolution spectra are used to derive the physical conditions and chemical abundances from both CELs and ORL, aiming at a redetermination of the \textit{adf}. The high-resolution spectra are used to analyze the internal kinematics of the nebula in emission lines of different excitation and ionization potential, aiming at obtaining a more detailed description of the morphokinematic and ionization structure of this object. With our results we provide relevant hints in our understanding of the formation and evolution of the nascent PN Vy\,2-2.

%--------------------------------------------------------------------
\section{Observations and data reduction}

\subsection{Intermediate-resolution optical spectroscopy}

Intermediate-resolution, long-slit spectra were obtained on Calar Alto Observatory (Almer{\'i}a, Spain) with the Calar Alto Faint Object Spectrograph (CAFOS) at the 2.2\,m telescope on 2016 August 24 (proposal ID: H16-2.2-024, PI: L.F. Miranda). The detector was a SITe CCD with 2048$\times$2048 pixels. We used grisms B-100 and R-100 to cover the spectral ranges 3200-6200 $\AA$ and 5800-9600 $\AA$, respectively, at a dispersion of $\sim$2.0 $\AA$ px$^{-1}$. The spatial scale in the detector is 0$\farcs$53\,pixel$^{-1}$. In Fig. \ref{fig:HST_caha} we show the position and orientation of the used long-slit (width of 1$\farcs$5). It was centered on the object, and oriented at PA = 35$^{\circ}$, hence covering the very bright core and the weak bipolar formation. The spectra were taken with exposure times of 1200, 300, 120, and 60\,s using the grism B-100, and of 1200, 300, 60, 30, and 20\,s using the grism R-100. The spectrophotometric standard star BD+28$^{\circ}$4211 was also observed for flux calibration. Sky was photometric during the observations and the seeing was $\sim$1$\farcs$5. 

Each individual spectrum was reduced following standard procedures in IRAF\footnote{IRAF is distributed by the National Optical Astronomy Observatories, which is operated the Association of Universities for Research in Astronomy, Inc., under contract to the National Science Foundation.} package. We note that the flux calibration failed at the very red edge of the CCD, coinciding with the [S\,{\sc iii}]$\lambda$9532 emission line; therefore the fluxes of this emission line are unreliable and will be not considered further in this paper. The use of very different exposure times allows us an optimal measurement of the strongest emission lines in the short exposure spectra, of very faint emission lines in deepest ones, and several measurements of emission lines with intermediate intensity. A gaussian or multigaussian (in the case of blended lines) line fit was used to measured the emission line fluxes. The individual flux measurements of each emission line in each spectrum were finally combined into a single mean value, weighted by the exposure time. The physical conditions and ionic abundances were derived using Pyneb \citep{lur15}.

\subsection{High-resolution optical and near-infrared spectroscopy}
\label{sec:carmen}

High-resolution optical and near-infrared \'echelle spectra were obtained on Calar Alto Observatory on 2016 July 12 with the Calar Alto high-Resolution search for M dwarfs with Exoearths making use of the Near-infrared and optical {\'E}chelle Spectrographs (CARMENES) (proposal ID: H16-3.5-0.36; PI: L.F. Miranda). The detectors for the optical and near-infrared spectra were an e2v CCD with 4096$\times$4096 pixels, and two Teledyne Hawaii arrays with 2048$\times$2048 pixels, respectively. CARMENES is coupled to the 3.5\,m telescope with two optical fibers (diameter=1$\farcs$5), one for the target, and one for sky. The wavelength coverage of the spectra is 5200-9600 $\AA$ and 9600-17100 $\AA$ for the optical and near-infrared ranges, respectively. The spectral resolution (FWHM) is $\simeq$2.4--3.2\,km\,s$^{-1}$ in the visible channel and $\simeq$4.3--5.4\,km\,s$^{-1}$ in the infrared one. The CARMENES fiber for the object was centered on Vy\,2-2 and covers the bright central region and the bipolar lobes (Fig\,1). Seeing was $\simeq$1$\farcs$5 and no flux calibration was carried out because we are interested in the internal kinematics of the object.

Five optical and near-infrared spectra were taken with an exposure time of 300\,s each. The spectra were reduced with the pipeline {\sc caracal} \citep{cab16}. The five individual visible and five near-infrared spectra were combined into one final spectrum at either wavelength range, making use of a statistical median, to increase the signal-to-noise ratio and to remove cosmic rays. The analysis of these final spectra was done using {\sc iraf} and the final errors were obtained propagating errors from each individual spectrum. We have also inspected the sky fiber to confirm that relevant sky emission lines do not exist when compared with those from Vy\,2-2. It should be noted that CARMENES uses vacuum wavelengths to carry out the wavelength calibration; consequently, to extract the information from the emission lines (e.g., radial velocity) we used the vacuum rest wavelengths as obtained from NIST\footnote{https://www.nist.gov/pml/atomic-spectra-database} database. Nevertheless, in the figures and tables of this paper we will use the more common air wavelengths.

\section{Results}

\begin{table*}
	\caption{Observed fluxes and intrinsic emission line intensities in units of $F$(H$\beta$)=100 and $I$(H$\beta$)=100, respectively, in Vy\,2-2 .} 
	\label{table:cafos}      % is used to refer this table in the text
	\centering                       % used for centering table
	\begin{tabular}{l c c c l c c c}        % centered columns (4 columns)
		\hline                 % inserts double horizontal lines
		\rule{0pt}{2ex}
		Emission line & $f(\lambda)$  & $F(\lambda)$ & $I(\lambda)$ & Emission line & $f(\lambda)$  & $F(\lambda)$ & $I(\lambda)$ \\    % table heading 
		\hline                        % inserts single horizontal line
		\rule{0pt}{2ex}

3705 He\,{\sc i} + H\,16	&	0.260	&	0.35	&	0.95 $\pm$ 0.21	&	5812 C\,{\sc iv}	&	$-$0.203	&	0.49	&	0.231 $\pm$ 0.011	\\
3712 H\,15	&	0.255	&	0.39	&	1.05 $\pm$ 0.18	&	5874 He\,{\sc i}	&	$-$0.216	&	57.7	&	25.76 $\pm$ 0.44	\\
3727 [O\,{\sc ii}]	&	0.250	&	4.57	&	12.06 $\pm$ 0.55	&	5958 O\,{\sc i} + Si\,{\sc ii}	&	$-$0.229	&	0.33	&	0.138 $\pm$ 0.019	\\
3750 H\,12	&	0.247	&	0.85	&	2.23 $\pm$ 0.31	&	5978 Si\,{\sc ii}	&	$-$0.231	&	0.22	&	0.095 $\pm$ 0.008	\\
3770 H\,11	&	0.244	&	1.01	&	2.60 $\pm$ 0.24	&	6045 O\,{\sc i}	&	$-$0.245	&	0.33	&	0.134 $\pm$ 0.019	\\
3798 H\,10	&	0.245	&	1.41	&	3.65 $\pm$ 0.25	&	6300 [O\,{\sc i}]	&	$-$0.285	&	3.54	&	1.215 $\pm$ 0.044	\\
3818 He\,{\sc i}	&	0.234	&	0.25	&	0.61 $\pm$ 0.08	&	6312 [S\,{\sc iii}]	&	$-$0.287	&	8.74	&	2.976 $\pm$ 0.041	\\
3835 H\,9	&	0.236	&	1.77	&	4.43 $\pm$ 0.24	&	6347 Si\,{\sc iii}	&	$-$0.292	&	0.46	&	0.153 $\pm$ 0.011	\\
3868 [Ne\,{\sc iii}]	&	0.228	&	19.3	&	46.85 $\pm$ 1.43	&	6363 [O\,{\sc i}]	&	$-$0.294	&	1.27	&	0.420 $\pm$ 0.011	\\
3889 He\,{\sc i} + H\,8	&	0.223	&	4.17	&	9.91 $\pm$ 1.35	&	6548 [N\,{\sc ii}]	&	$-$0.321	&	22.6	&	6.77 $\pm$ 0.18	\\
3967 H$\epsilon$ + [Ne\,{\sc iii}]	&	0.203	&	19.3	&	42.60 $\pm$ 0.67	&	6563 H$\alpha$	&	$-$0.323	&	956.2	&	285.0 $\pm$ 9.3	\\
3997 [F\,{\sc iv}] + N\,{\sc ii}	&	0.195	&	0.07	&	0.141 $\pm$ 0.007	&	6583 [N\,{\sc ii}]	&	$-$0.326	&	62.3	&	18.3 $\pm$ 0.6	\\
4024 He\,{\sc i}	&	0.189	&	1.05	&	2.19 $\pm$ 0.20	&	6678 He\,{\sc i}	&	$-$0.338	&	17.3	&	4.87 $\pm$ 0.07	\\
4040 N\,{\sc ii}	&	0.189	&	0.05	&	0.105 $\pm$ 0.008	&	6716 [S\,{\sc ii}]	&	$-$0.343	&	1.11	&	0.305 $\pm$ 0.012	\\
4045 N\,{\sc ii}	&	0.189	&	0.03	&	0.061 $\pm$ 0.008	&	6731 [S\,{\sc ii}]	&	$-$0.345	&	2.06	&	0.563 $\pm$ 0.012	\\
4060 [F\,{\sc iv}]	&	0.180	&	0.17	&	0.350 $\pm$ 0.024	&	7000 O\,{\sc i}	&	$-$0.376	&	0.52	&	0.1270 $\pm$ 0.0038	\\
4068 [S\,{\sc ii}]	&	0.180	&	1.31	&	2.64 $\pm$ 0.13	&	7065 He\,{\sc i}	&	$-$0.383	&	48.7	&	11.51 $\pm$ 0.30	\\
4069 [O\,{\sc ii}] + [O\,{\sc ii}]	&	0.178	&	0.55	&	1.110 $\pm$ 0.033	&	7135 [Ar\,{\sc iii}]	&	$-$0.390	&	97.6	&	22.5 $\pm$ 1.6	\\
4077 [O\,{\sc ii}] + [O\,{\sc ii}]	&	0.178	&	0.29	&	0.59 $\pm$ 0.05	&	7155 [Fe\,{\sc ii}]	&	$-$0.393	&	1.36	&	0.309 $\pm$ 0.012	\\
4089 O\,{\sc ii}	&	0.177	&	0.15	&	0.290 $\pm$ 0.024	&	7231 [C\,{\sc ii}]	&	$-$0.401	&	0.86	&	0.191 $\pm$ 0.007	\\
4102 H$\delta$	&	0.172	&	9.82	&	19.23 $\pm$ 0.33	&	7254 O\,{\sc i}	&	$-$0.403	&	0.77	&	0.169 $\pm$ 0.007	\\
4144 He\,{\sc i}	&	0.163	&	0.28	&	0.527 $\pm$ 0.037	&	7281 He\,{\sc i}	&	$-$0.406	&	4.88	&	1.057 $\pm$ 0.020	\\
4266 C\,{\sc ii}	&	0.141	&	0.16	&	0.269 $\pm$ 0.021	&	7298 He\,{\sc i}	&	$-$0.407	&	0.33	&	0.0720 $\pm$ 0.0040	\\
4340 H$\gamma$	&	0.129	&	23.6	&	39.2 $\pm$ 0.5	&	7319 [O\,{\sc ii}]	&	$-$0.410	&	48.4	&	10.31 $\pm$ 0.21	\\
4363 [O\,{\sc iii}]	&	0.124	&	4.81	&	7.85 $\pm$ 0.49	&	7330 [O\,{\sc ii}]	&	$-$0.411	&	45.8	&	9.73 $\pm$ 0.20	\\
4388 He\,{\sc i}	&	0.117 	&	0.48	&	0.76 $\pm$ 0.19 	&	7378 [Ni\,{\sc ii}]	&	$-$0.416	&	0.79	&	0.1644 $\pm$ 0.0032	\\
4471 He\,{\sc i}	&	0.095	&	4.25	&	6.20 $\pm$ 0.05	&	7387 Fe\,{\sc ii}	&	$-$0.416	&	0.38	&	0.0790 $\pm$ 0.0031	\\
4591 O\,{\sc ii}	&	0.065	&	0.07	&	0.0852 $\pm$ 0.0033	&	7410 Si\,{\sc ii}	&	$-$0.420	&	0.38	&	0.0778 $\pm$ 0.0042	\\
4596 O\,{\sc ii}	&	0.065	&	0.06	&	0.0772 $\pm$ 0.0033	&	7452 [Fe\,{\sc ii}]	&	$-$0.423	&	0.26	&	0.0523 $\pm$ 0.0018	\\
4631 N\,{\sc ii}	&	0.053	&	0.22	&	0.271 $\pm$ 0.018	&	7466 N\,{\sc i}	&	$-$0.424	&	0.17	&	0.0346 $\pm$ 0.0018	\\
4640 O\,{\sc ii} + O\,{\sc ii} + N\,{\sc ii}	&	0.052	&	0.58	&	0.718 $\pm$ 0.019	&	7498 He\,{\sc i}	&	$-$0.427	&	0.37	&	0.0730 $\pm$ 0.0019	\\
4649 O\,{\sc ii} + O\,{\sc ii} + C\,{\sc iii}	&	0.052	&	0.80	&	0.985 $\pm$ 0.019	&	7751 [Ar\,{\sc iii}]	&	$-$0.451	&	27.1	&	4.95 $\pm$ 0.11	\\
4658 C\,{\sc iv} + Fe\,{\sc iii}	&	0.050	&	1.14	&	1.404 $\pm$ 0.026	&	7816 He\,{\sc i}	&	$-$0.457	&	0.97	&	0.174 $\pm$ 0.007	\\
4686 He\,{\sc ii}	&	0.046	&	1.51	&	1.824 $\pm$ 0.031	&	7888 Fe\,{\sc i}	&	$-$0.463	&	0.26	&	0.0455 $\pm$ 0.0013	\\
4701 Fe\,{\sc iii}	&	0.036	&	0.63	&	0.744 $\pm$ 0.030	&	7998 Fe\,{\sc i}	&	$-$0.475	&	0.13	&	0.0209 $\pm$ 0.0014	\\
4712 [Ar\,{\sc iv}]	&	0.033	&	1.06	&	1.239 $\pm$ 0.025	&	8045 C\,{\sc ii}	&	$-$0.477	&	0.44	&	0.0717 $\pm$ 0.0024	\\
4733 [Fe\,{\sc iii}]	&	0.031	&	0.22	&	0.247 $\pm$ 0.028	&	8392 P\,20	&	$-$0.509	&	3.62	&	0.53 $\pm$ 0.05	\\
4740 [Ar\,{\sc iv}]	&	0.029	&	0.22	&	0.247 $\pm$ 0.045	&	8413 P\,19	&	$-$0.512	&	3.96	&	0.57 $\pm$ 0.05	\\
4753 [Fe\,{\sc iii}]	&	0.027	&	0.25	&	0.276 $\pm$ 0.036	&	8438 P\,18	&	$-$0.516	&	4.73	&	0.67 $\pm$ 0.05	\\
4768 [Fe\,{\sc iii}]	&	0.024	&	0.15	&	0.172 $\pm$ 0.007	&	8446 O\,{\sc i}	&	$-$0.517	&	11.4	&	1.62 $\pm$ 0.05	\\
4813 [Fe\,{\sc iii}]	&	0.012	&	0.07	&	0.0722 $\pm$ 0.0029	&	8467 P\,17	&	$-$0.521	&	5.33	&	0.74 $\pm$ 0.05	\\
4861 H$\beta$	&	0.000	&	100.0	&	100.0 $\pm$ 2.5	&	8502 P\,16	&	$-$0.526	&	6.73	&	0.92 $\pm$ 0.05	\\
4880 [Fe\,{\sc iii}]	&	$-$0.005	&	0.41	&	0.41 $\pm$ 0.42	&	8528 He\,{\sc i}	&	$-$0.530	&	0.46	&	0.062 $\pm$ 0.030	\\
4922 He\,{\sc i}	&	$-$0.014	&	1.71	&	1.646 $\pm$ 0.041	&	8545 P\,15	&	$-$0.532	&	6.92	&	0.93 $\pm$ 0.05	\\
4959 [O\,{\sc iii}]	&	$-$0.023	&	235.9	&	219.7 $\pm$ 18.0	&	8579 Cl\,{\sc ii}	&	$-$0.537	&	1.42	&	0.19 $\pm$ 0.06	\\
5007 [O\,{\sc iii}]	&	$-$0.034	&	711.2	&	635.3 $\pm$ 8.5	&	8598 P\,14	&	$-$0.540	&	8.15	&	1.06 $\pm$ 0.05	\\
5159 Fe\,{\sc i} + C\,{\sc i}	&	$-$0.064	&	0.25	&	0.199 $\pm$ 0.042	&	8617 [Fe\,{\sc ii}]	&	$-$0.543	&	1.44	&	0.185 $\pm$ 0.027	\\
5192 [Ar\,{\sc iii}]	&	$-$0.073	&	0.25	&	0.195 $\pm$ 0.019	&	8647 He\,{\sc i}	&	$-$0.547	&	0.71	&	0.090 $\pm$ 0.029	\\
5199 N\,{\sc i}	&	$-$0.074	&	0.15	&	0.113 $\pm$ 0.017	&	8665 P\,13	&	$-$0.550	&	9.51	&	1.19 $\pm$ 0.05	\\
5270 [Fe\,{\sc iii}]	&	$-$0.089	&	1.17	&	0.848 $\pm$ 0.046	&	8750 P\,12	&	$-$0.562	&	11.5	&	1.38 $\pm$ 0.05	\\
5411 He\,{\sc ii}	&	$-$0.118	&	0.26	&	0.170 $\pm$ 0.022	&	8863 P\,11	&	$-$0.578	&	14.9	&	1.674 $\pm$ 0.030	\\
5517 [Cl\,{\sc iii}]	&	$-$0.139	&	0.20	&	0.119 $\pm$ 0.014	&	8995 He\,{\sc i}	&	$-$0.596	&	1.19	&	0.125 $\pm$ 0.017	\\
5537 [Cl\,{\sc iii}]	&	$-$0.143	&	0.51	&	0.298 $\pm$ 0.024	&	9015 P\,10	&	$-$0.599	&	21.8	&	2.26 $\pm$ 0.17	\\
5666 N\,{\sc ii}	&	$-$0.147	&	0.12	&	0.0688 $\pm$ 0.0019	&	9069 [S\,{\sc iii}]	&	$-$0.606	&	441.6	&	44.60 $\pm$ 1.44	\\
5676 N\,{\sc ii} + 5679 N\,{\sc ii}	&	$-$0.147	&	0.05	&	0.1497 $\pm$ 0.0024	&	9209 He\,{\sc i}	&	$-$0.612	&	5.71	&	0.56 $\pm$ 0.05	\\
5755 [N\,{\sc ii}]	&	$-$0.191	&	5.85	&	2.872 $\pm$ 0.037	&	9229 P\,9	&	$-$0.612	&	80.8	&	7.98 $\pm$ 0.10	\\
5801 C\,{\sc iv}	&	$-$0.201	&	0.54	&	0.253 $\pm$ 0.006	&	9462 He\,{\sc i}	&	$-$0.618	&	61.0	&	5.89 $\pm$ 0.14	\\

            \hline
		\rule{0pt}{1ex}			
		log $F_{H\beta}$\,(erg\,cm$^{-2}$\,s$^{-1}$) & & & $-$11.649 $\pm$ 0.003 \\
		\rule{0pt}{1ex}	
		c(H$\beta$) & & & 1.66 $\pm$ 0.03\\
		  
		\hline                                   %inserts single line
	\end{tabular}
\end{table*}

\subsection{Nebular emission lines}

Table\,\ref{table:cafos} lists the nebular emission lines identified in the intermediate-resolution spectra of Vy\,2-2. From the H$\alpha$ and H$\beta$ emission line fluxes, a theoretical H$\alpha$/H$\beta$=2.85 intensity ratio, and the interstellar extinction law of \citep{sea79}, we obtain a logarithmic extinction coefficient at H$\beta$, $c$H($\beta$)=1.66$\pm$0.03. This value agrees with those reported by \cite{wes05} and \cite{ark17} but differs significantly from the value of $0.66 \pm 0.06$ reported by \cite{reb22}. However, the H$\alpha$/H$\beta$ ratio given by \cite{reb22} is consistent with a value of $c$(H$\beta$) $\simeq 1.61$, in good agreement with ours and previous values. Therefore, a typographical error in their Table 3 seems to be a possible explanation of such difference.

Table\,\ref{table:cafos} also shows the intrinsic emission line intensities using the interstellar extinction law of \cite{sea79}, and the observed H$\beta$ flux. In this paper we do not show the nebular spectrum of Vy\,2-2 because similar ones have already been published elsewhere \citep[e.g.][]{ark17}. However, a detailed inspection of our spectra allowed us to identified for the first time several ORLs and CELs, some of which are key to determine some physical conditions in Vy\,2-2. These lines will be individually mentioned, and relevant parts of the spectra shown later in this section.

\begin{figure}
	\centering
	\includegraphics[width=8.4cm]{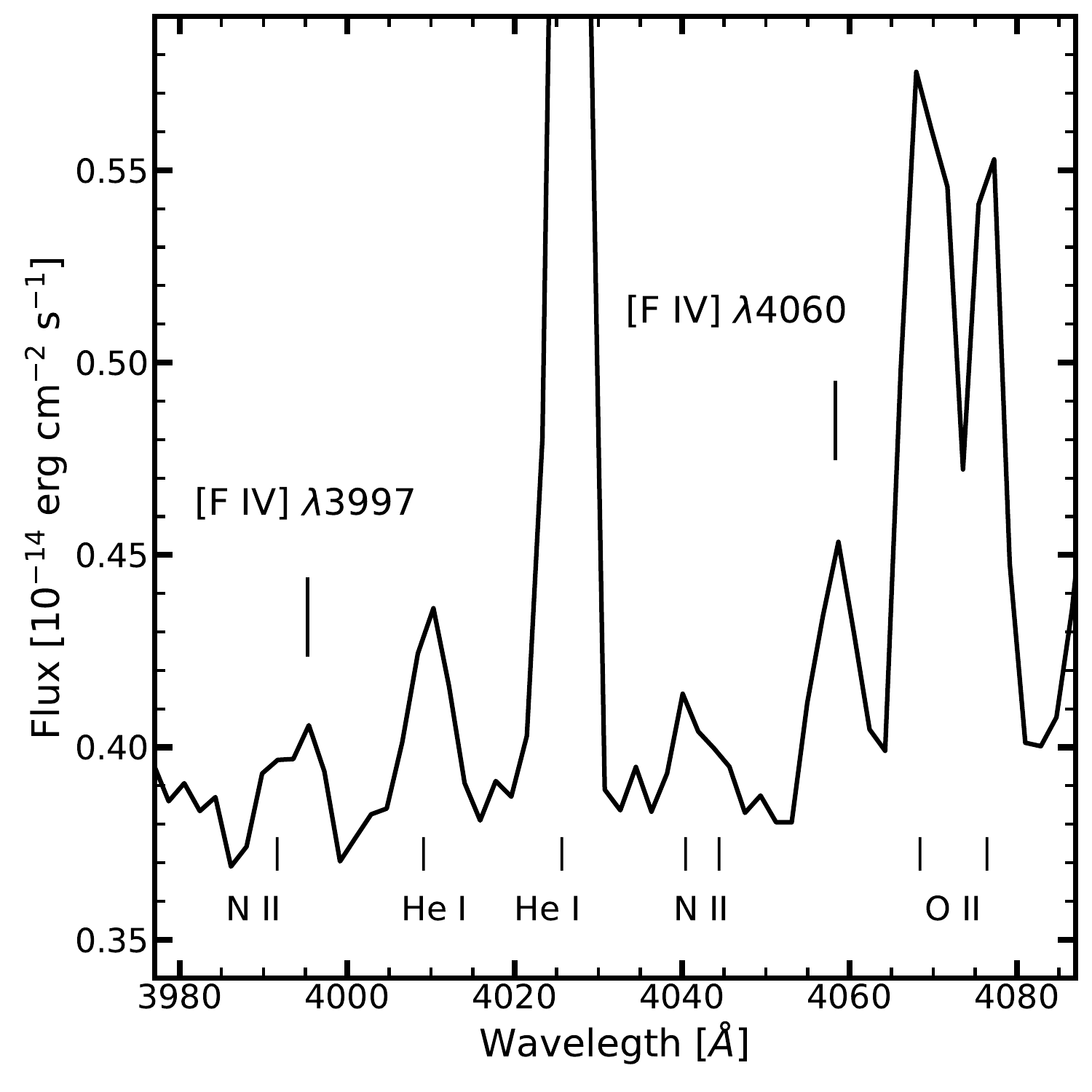}
	\caption{The 3980-4085\,$\AA$ region of deepest spectrum of Vy\,2-2. The possible [F\,{\sc iv}] $\lambda$3997,4060 emission lines as well as other recombination emission lines are marked.}
	\label{fig:Fluorine}
\end{figure}

\subsection{Fluorine in Vy\,2-2?}

In Fig.\,\ref{fig:Fluorine} we show the 3980-4085\,$\AA$ region as observed in our deepest blue spectrum of Vy\,2-2. We have identified two emission lines which may be the forbidden ones of the triply ionized doublet of fluorine, namely, [F\,{\sc iv}]$\lambda$3997,4060. They are clearly identified only in this spectrum. In Table,\,\ref{table:cafos} we include their dereddened emission line intensities and we note that the $\lambda$3997 line is blended with the N\,{\sc ii}\,$\lambda$3991 one (Fig.\,\ref{fig:Fluorine}). Using a two-gaussian line fit to separate these two emission lines we obtained $I$($\lambda$4060/$\lambda$3997) $\simeq$2.5, which is similar to the theoretical value of $\simeq$2.8 \citep{zha05, ots08}. The possible detection of [F\,{\sc iv}] could suggest the presence of fluorine in lower ionization stages, such as [F\,{\sc ii}] or [F\,{\sc iii}]. However, we do not detect the [F\,{\sc ii}]$\lambda$4789 line, which has also not been detected in other PNe where [F\,{\sc iv}]$\lambda$$\lambda$3997,4060 have been reported \citep{zha05, ots08}. Nevertheless, whether this line should be detectable depends on the nebular conditions and the relevant atomic parameters, and its non-detection does not independently support our proposed identifications. If our identifications are correct, they would be the first report of F in Vy\,2-2. This would be particularly interesting given the possible connection between fluorine production and binary-star evolution \citep{ots08}, in the context of the other properties of Vy\,2-2 that suggest the presence of a close binary system (Sect.\,\ref{sec:binary}).

\subsection{Physical conditions, chemical abundances, and adf}\label{sec:physical}

We used the line intensity ratios between the CELs [O\,{\sc iii}]$\lambda$(4959+5007)/4363, [O\,{\sc ii}]$\lambda$(3726+3729)/(7320+7330), and [Ar\,{\sc iii}]$\lambda$(7136+7751)/5192 to derive values of electron temperature ($T_{\rm e}$), and those of [S\,{\sc ii}]$\lambda$6716/6731 and [Cl\,{\sc iii}]$\lambda$5517/5537 to derive values of electron density ($N_{\rm e}$). We did not use the [N\,{\sc ii}] $\lambda$(6548+6584)/5755 line intensity ratio to derive $T_{\rm e}$ because this diagnostic may be affected by both collisional de-excitation and recombination contributions to the auroral $\lambda$5755 line \citep{liu00, gar22}. For this reason, we consider the resulting $T_{\rm e}$([N\,{\sc ii}]) to be unreliable and do not use it in our analysis, following the same approach as \cite{reb22}. We found that the ionized gas seems to be composed of several zones with different physical conditions ($T_{\rm e}$, $N_{\rm e}$). In Table\,\ref{table:zones} we report these zones as Zone\,1, 2, and 3, together with their corresponding values of $T_{\rm e}$ and $N_{\rm e}$. \cite{reb22} also identified three different zones in the ionized gas of Vy\,2-2 with different physical conditions, although their zones present some differences with respect to ours. Nevertheless, we agree with \cite{reb22} in the existence of a zone (Zone\,3) in which Vy\,2-2 clearly presents significantly higher $T_{\rm e}$ and $N_{\rm e}$ than in Zones\,1 and 2.

\begin{table*}
	\caption{Zones in Vy\,2-2, their $T_{\rm e}$ and $N_{\rm e}$, and mean ionic abundances relative to H$^{+}$ obtained from CELs.} % title of Table
	\label{table:zones}      % is used to refer this table in the text
	\centering                          % used for centering table
	\begin{tabular}{l l l l l}        % centered columns (4 columns)
		
		\hline                 % inserts double horizontal lines
		%\rule{0pt}{2.3ex}
		Zone & $T_{\rm e}$ [K] & $N_{\rm e}$ [cm$^{-3}$] & Ion & Ionic abundance \\
		\hline
		%\rule{0pt}{3ex}
		1 & [Ar\,{\sc iii}] 10700$^{+700}_{-650}$ & [S\,{\sc ii}] 7250$^{+3900}_{-2050}$ &  \\
        \vspace{0.1cm}
		%\rule{0pt}{3ex}		
		& & & S$^{+}$  & 3.83$^{+1.15}_{-0.61}$ $\times$ 10$^{-8}$ \\	
        \vspace{0.1cm}
		%rule{0pt}{3ex}							   
		& & & O$^{0}$ & 1.87$^{+0.52}_{-0.40}$ $\times$ 10$^{-6}$ \\	
		%\rule{0pt}{3ex}
        \vspace{0.1cm}
		& & & N$^{0}$   & 5.86$^{+2.38}_{-1.27}$ $\times$ 10$^{-7}$ \\
        \vspace{0.1cm}
		%\rule{0pt}{3ex}
		& & & N$^{+}$  & 3.26$^{+0.56}_{-0.47}$ $\times$ 10$^{-6}$ \\
        \vspace{0.1cm}
		%\rule{0pt}{3ex}
		& & & Ar$^{++}$  & 1.74$\pm$0.26 $\times$ 10$^{-6}$ \\	
		%\rule{0pt}{3ex}
		2 	& [O\,{\sc iii}] 12400$^{+450}_{-450}$ & [S\,{\sc ii}] 7250$^{+3900}_{-2050}$ &  \\
		%\rule{0pt}{3ex}
		& & & O$^{++}$ & 1.12$\pm$0.19 $\times$ 10$^{-4}$ \\
		%\rule{0pt}{3ex}
        \vspace{0.1cm}
		& & & S$^{++}$ & 2.96$^{+0.41}_{-0.35}$ $\times$ 10$^{-6}$ \\	
		%\rule{0pt}{3ex}
		3 	& [O\,{\sc ii}] 16800$^{+1600}_{-1300}$ & [Cl\,{\sc iii}] 63400$^{+}_{-38600}$\tablefootmark{a}  \\
            %\rule{0pt}{3ex}
            \vspace{0.1cm}
		& & & O$^{+}$ & 8.50$^{+2.52}_{-3.27}$ $\times$ 10$^{-6}$ \\	
		%\rule{0pt}{3ex}
        \vspace{0.1cm}
		& & & Ar$^{+++}$ & 1.20$^{+0.24}_{-0.24}$ $\times$ 10$^{-8}$ \\
		%\rule{0pt}{3ex}
        \vspace{0.1cm}
		& & & Ne$^{++}$ & 8.59$^{+2.03}_{-1.78}$ $\times$ 10$^{-6}$ \\			
		%\rule{0pt}{3ex}
        \vspace{0.1cm}
		& & & Cl$^{++}$ & 1.82$^{+0.98}_{-0.44}$ $\times$ 10$^{-8}$ \\		
		%\rule{0pt}{3ex}
        \vspace{0.1cm}
		& & & Cl$^{+++}$  & 1.88$^{+0.26}_{-0.26}$ $\times$ 10$^{-9}$ \\
         %   \rule{0pt}{3ex}
         \vspace{0.1cm}
            & & & He$^{+}$\tablefootmark{b}& 11.69$^{+2.50}_{-1.59}$ $\times$ 10$^{-2}$ \\
		%\rule{0pt}{3ex}
		& & & He$^{++}$\tablefootmark{b} & 1.47$\pm$0.04 $\times$ 10$^{-3}$ \\

		\hline                                   %inserts single line
	\end{tabular}
	\tablefoot{
  \tablefootmark{a}{The upper limit for the ionic abundances in Zone\,3 used the mean value of $N_{\rm e}$ and the error of the dereddened emission line. This is because the [Cl\,{\sc iii}] diagnostic ratio is temperature-sensitive and gets unstable at the upper limit of $T_{\rm e}$([O\,{\sc ii}]).}
   \tablefootmark{b}{Averaged value from the three zones. For Zone\,3, a maximum value of $N_{e}$=\,63400\,cm$^{-3}$ was used.}
  }
  
\end{table*}

\begin{table}
	\caption{Elemental abundances (12+log(X/H) scale) in Vy\,2-2 derived from CELs obtained in this work, by \cite{wes05} (WLB05) and \cite{reb22} (REP22), and in the Sun \citep{asp09}. } % title of Table
	\label{table:total}      % is used to refer this table in the text
	\centering                          % used for centering table
	\begin{tabular}{c c c c c}        % centered columns (4 columns)
		\hline
		%\rule{0pt}{2.5ex}                 % inserts double horizontal lines
		& WLB05 & REP22  & This work & The Sun\\    % table heading 
		\hline                        % inserts single horizontal line
		%\rule{0pt}{3ex}
		He  & 11.03 & 11.06$\pm$0.01 & 11.07$^{+0.08}_{-0.06}$ & 10.93$\pm$0.01 \\
		%\rule{0pt}{3ex}
		O & 7.98 & 8.46$^{+0.12}_{-0.05}$ & 8.09$^{+0.07}_{-0.09}$ & 8.69$\pm$0.05 \\
		%\rule{0pt}{3ex}
		N & 7.83 & 7.74$^{+0.13}_{-0.09}$ & 7.70$^{+0.03}_{-0.06}$ & 7.83$\pm$0.05 \\
		%\rule{0pt}{3ex}
		Ne & 7.24 & 7.58$^{+0.11}_{-0.05}$ & 7.36$^{+0.24}_{-0.21}$ & 7.93$\pm$0.10 \\
		%\rule{0pt}{3ex}
		S & 6.65 & 7.31$\pm$0.05 & 6.71$^{+0.04}_{-0.03}$ & 7.12$\pm$0.03  \\
		%\rule{0pt}{3ex}
		Ar  & 5.88 & 6.48$\pm$0.03 & 6.24$^{+0.06}_{-0.07}$ & 6.40$\pm$0.13 \\
		%\rule{0pt}{3ex}
		Cl & - & 4.97$^{+0.23}_{-0.11}$ & 4.26$^{+0.19}_{-0.12}$ & 5.50$\pm$0.30 \\
		%\rule{0pt}{3ex}
		Fe & - & 6.35$^{+0.08}_{-0.06}$ &  & 7.50$\pm$0.04 \\
				
		\hline                                   %inserts single line
	\end{tabular}
\end{table}

Table\,\ref{table:zones} also shows the ionic abundances. These values were calculated taking into account the particular excitation conditions and ionization potential of each ion, as listed in Table\,\ref{table:zones}. In general, our ionic abundances are similar to (or in between of) those reported by \cite{wes05} and \cite{reb22}. We note that the ionic abundance of F has not been obtained because it requires knowing of Ne$^{3+}$, which is not available. For the elemental abundances, we use the ionization correction factors by \cite{kin94} and they are reported in Table \ref{table:total}, where we compare our values with those by \cite{wes05}, \cite{reb22}, and in the Sun \citep{asp09}. Our results are compatible with (or in between of) those reported by \cite{wes05} and \cite{reb22}, and we note that the abundances of Vy 2-2 are subsolar. 

Physical conditions have been also obtained from ORLs (see Appendix\,\ref{sec:orl_physical} for details about line identification). From the O\,{\sc ii}\,$\lambda$4649.13/4661.63 and $\lambda$4075.86/4070 line intensity ratios we obtain $T_{\rm e}$(O\,{\sc ii})=6309$^{+769}_{-686}$\,K, and $N_{\rm e}$(O\,{\sc ii})=4467$^{+781}_{-664}$ cm$^{-3}$, respectively. We use Pyneb to determine the individual line abundances based on these physical conditions. The obtained values are listed in Table \ref{table:IA_ORLs} and are in general agreement with those reported by \cite{wes05}. In particular, they reported O$^{++}_{\rm ORL}$/H$^{+}$$\simeq$10.86$\times$10$^{-4}$, which is comparable to $\simeq$15.23$\times$10$^{-4}$ obtained by us. \cite{reb22} found a different value of O$^{++}_{\rm ORL}$/H$^{+}$$\simeq$5.50$\times$10$^{-4}$. To investigate whether this discrepancy could be related to the adopted electron density, we recalculated O$^{++}_{\rm ORL}$/H$^{+}$ using the $N{\rm e}$(O\,{\sc ii})=$19900\pm4300$ cm$^{-3}$ reported by \cite{reb22}, obtaining O$^{++}_{\rm ORL}$/H$^{+}\simeq16.0 \times$10$^{-4}$. This result is similar to our original value and therefore indicates that the different electron densities obtained in the two studies cannot account for the differences. We also tested the effect of different electron temperatures and found that this has only a minor effect on the derived abundance. However, when we adopt the O\,{\sc ii} recombination-line intensities reported by \cite{reb22}, while keeping our $T_{\rm e}$ and $N_{\rm e}$ values, we obtain O$^{++}_{\rm ORL}$/H$^+$$\simeq$6.3$\times$10$^{-4}$, which is similar to their reported value. This indicates that the different line intensities obtained in the two studies are primarily responsible for the discrepancy in the derived O$^{++}$/H$^+$ abundances. The possible effects of line blending and stellar contamination on the identification and measurement of these lines are discussed in Appendix\,\ref{sec:orl_physical}.

To determine the N/O abundance ratio from ORLs, we used the expression N/O\,=\,ICF(N$^{++}$/O$^{++}$)$\times$(N$^{++}_{\rm ORL}$/O$^{++}_{\rm ORL}$) as proposed by \cite{del14} and their equations\,14, 15, and 16. We obtained ICF(N$^{++}$/O$^{++}$)=0.99$^{+0.30}_{-0.15}$ and, hence, N/O=0.23$^{+0.07}_{-0.09}$. This value is somewhat lower than that of 0.41$\pm$0.09 derived from CELs (Table\,\ref{table:total}). However, our values are even more discrepant from N/O$\simeq$0.66 obtained by \cite{wes05}, which was based on ORLs. To further understand the origin of this difference, we recalculated the N$^{++}$/H$^{+}$ abundance using the N\,{\sc iii}$\lambda$4630 recombination line, also employed by \cite{wes05}. We derive N$^{++}$/H$^{+}$ = (13.6$\pm$0.8)$\times$10$^{-4}$ and N/O$\simeq$0.88, in good agreement with their values of N$^{++}$/H$^{+}$$\simeq$11.65$\times$10$^{-4}$ and N/O$\simeq$0.66. This agreement indicates that the differences between the two analyses is mainly due to the different sets of recombination lines adopted: \cite{wes05} used N$^{++}$ lines that may be of stellar origin (see Sec.\,\ref{sec:wels} for more details), whereas we used nebular N$^{+}$ recombination lines, reported here for the first time (see Table\,\ref{table:IA_ORLs}).

The C/O abundance ratio was calculated with the expression C/O\,=\,ICF(C$^{++}$/O$^{++}$)$\times$(C$^{++}_{\rm ORL}$/O$^{++}_{\rm ORL}$) as proposed by \cite{del14}. From Fig.\,14 in \cite{del14}, we obtained log$_{10}$ICF(C$^{++}$/O$^{++}$)=0.0$^{+0.2}_{-0.1}$ and, hence, C/O=0.15$^{+0.09}_{-0.04}$. This value is compatible within the errors with the value of $\simeq$0.26 obtained by \cite{wes05}.

The ionic abundance reported in Tables \ref{table:zones} and \ref{table:IA_ORLs} yield an $adf$(O$^{2+}$)=13.6$\pm$2.6. This value is similar to $\simeq$11.8 reported by \cite{wes05}, but is very different from 4.30$^{+1.00}_{-1.16}$ reported by \cite{reb22}, providing support to the idea of an unusually high  $adf$(O$^{2+}$) in Vy\,2-2. Finally, we could not obtain $adf$(C$^{2+}$) and $adf$(N$^{2+}$) because C$^{++}$/H$^{+}$ and N$^{++}$/H$^{+}$ were obtained only for ORLs, and not for CELs.

\citet{wes18} found an anti-correlation between nebular density and abundance discrepancy factor, with most high-density objects having low abundance discrepancies. Thus, Vy\,2-2 occupies an unusual region of the $N_{\rm e}$--$adf$(O$^{2+}$) diagram presented in their Figure 16, combining a high $adf$(O$^{2+}$) with an exceptionally high $N_{\rm e}$([Cl\,{\sc iii}]). We also note that under very high electron density conditions, standard plasma diagnostics based on CELs in compact PNe may become uncertain, since transitions with relatively low critical densities can be significantly affected by collisionally de-excitation in the densest regions of the nebula. Therefore, the integrated CEL emission may not accurately trace the total chemical abundances of these emission lines.

\begin{table}
	\caption{Ionic abundances relative to H$^{+}$ from ORLs.} % title of Table
	\label{table:IA_ORLs}      % is used to refer this table in the text
	\centering                          % used for centering table
	\begin{tabular}{l c c c c}        % centered columns (4 columns)
		
		\hline                 % inserts double horizontal lines
		%\rule{0pt}{2.3ex}
		  Line &   Ion & [$\times$ 10 $^{-4}$] \tablefootmark{a}  \\
		\hline
		%\rule{0pt}{3ex}
		     & O$^{++}$/H$^{+}$ \\
		%\rule{0pt}{3ex}
		  $\lambda$4638.86 & & 18.07$\pm$4.90  \\
		%\rule{0pt}{3ex}
		  $\lambda$4641.81 & & 12.65$\pm$3.42 \\		
		%\rule{0pt}{3ex}
		  $\lambda$4649.13 & & 10.01$\pm$0.36  \\
		%\rule{0pt}{3ex}
		  $\lambda$4650.84 & & 17.33$\pm$1.03  \\
		%\rule{0pt}{3ex}
		   & & \textbf{15.23$\pm$1.27}\tablefootmark{b}   \\
		%\rule{0pt}{3ex}
		   & C$^{++}$/H$^{+}$ \\
		%\rule{0pt}{3ex}
		  $\lambda$4267.0 & &\textbf{2.34$\pm$0.25}\tablefootmark{b}   \\
		%\rule{0pt}{3ex}
		   & N$^{++}$/H$^{+}$  \\
		%\rule{0pt}{3ex}
		  $\lambda$5666.63 & & 3.74$\pm$0.12 \\		
		%\rule{0pt}{3ex}
		  $\lambda$5676.02 & & 3.28$\pm$0.23 \\
		%\rule{0pt}{3ex}
		  $\lambda$5679.56 & & 3.36$\pm$0.05 \\
		%\rule{0pt}{3ex}
		   & & \textbf{3.46$\pm$0.27}\tablefootmark{b}   \\
		\hline
		                            
	\end{tabular}
	\tablefoot{
  \tablefootmark{a}{The errors are the upper and lower values obtained using the maximum and minimum values of each of the variables involved ($N_{\rm e}$, $T_{\rm e}$, and $I(\lambda)$).}
  \tablefootmark{b}{Final adopted ionic abundances. For O\,{\sc ii} and N\,{\sc ii}, the adopted values are the average of the individual line abundances.}}
  	
\end{table}

\subsection{Stellar emission lines}\label{central_star}
\label{sec:wels}

\begin{figure*}
	\centering
	\includegraphics[width=8cm]{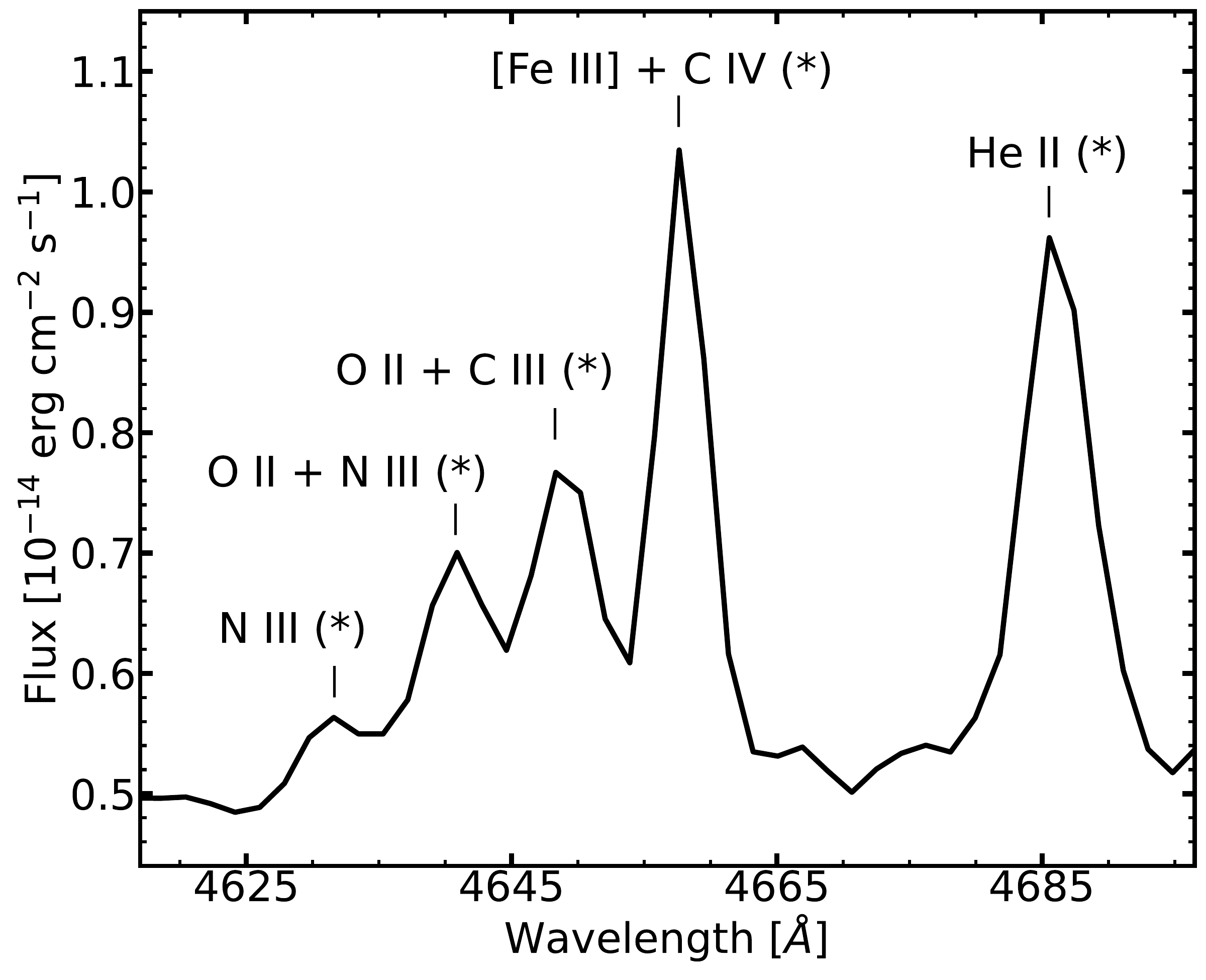}
	\hspace{0.25cm}
	\includegraphics[width=8.4cm]{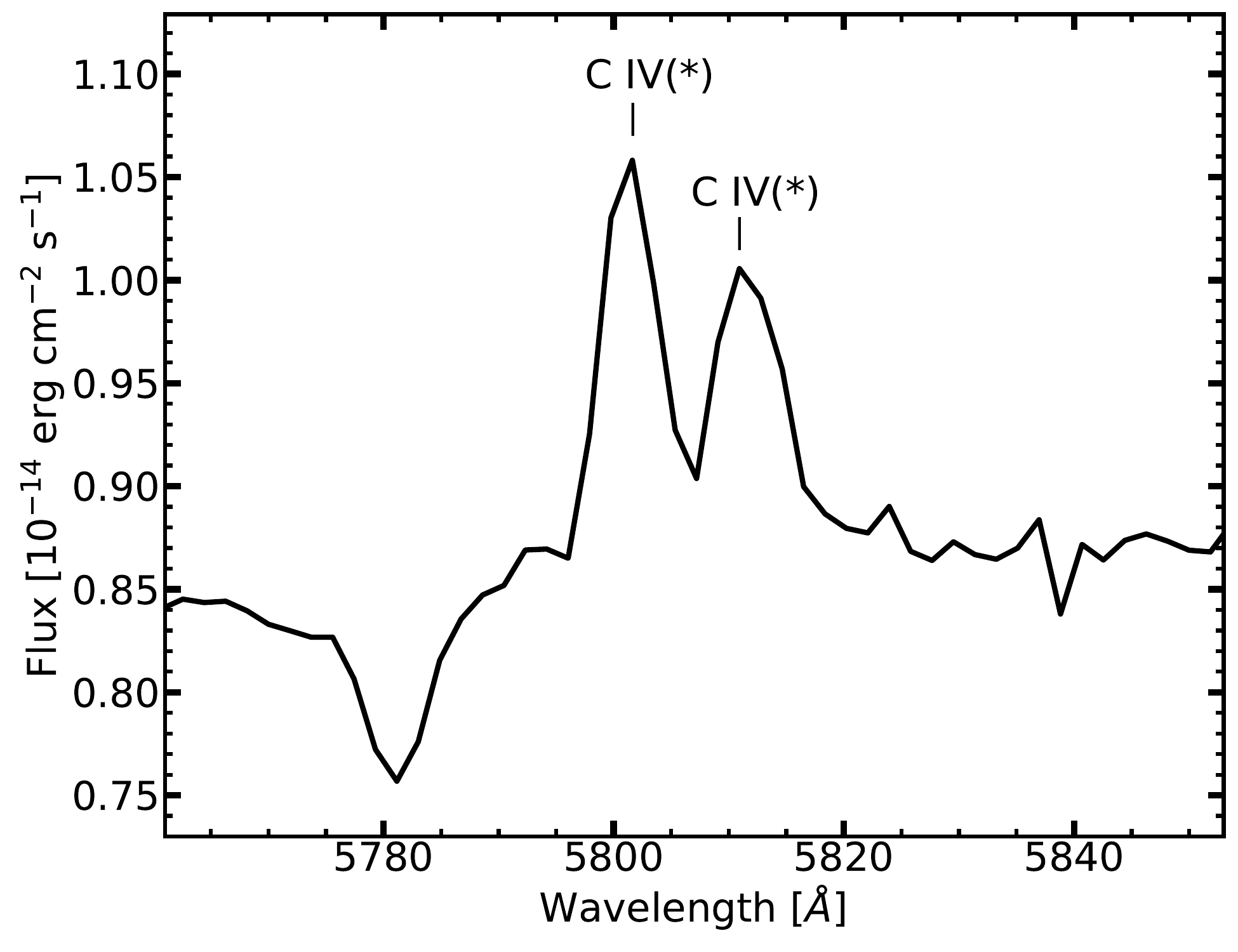}
	\caption{Weak emission lines (\textit{wels}) (*) in the optical blue spectrum of Vy\,2-2.}
	\label{fig:wels_C4}
\end{figure*}

The possible presence of stellar emission lines in Vy\,2-2 was already suggested by \cite{ark17}, and \cite{reb22} associated the He\,{\sc ii} $\lambda$4686 recombination line to the central star of Vy\,2-2. Fig.\,\ref{fig:wels_C4} presents our deepest blue spectrum in which the stellar emission lines are clearly observed and identified following \cite{mar93} and the nebular emission lines observed in the spectrum presented by \cite{reb22}. Table \ref{table:wels} lists the intrinsic line intensities of the stellar emission lines obtained with the same $c$(H$\beta$) as deduced for the nebula (Table\,\ref{table:cafos}).

The detection of the doublet of C\,{\sc iv}$\lambda\lambda$5801,5812 emission lines in our spectra is clear (Fig.\,\ref{fig:wels_C4}) and the observed value of $I(\lambda5801)/I(\lambda5812)$ $\simeq$\,4/3 is in agreement with the theoretical ratio \citep{moo45}. Besides, if we assume an 
$I(\lambda4658)/I(\lambda5812)$ theoretical ratio of 5/4, the intensity of C\,{\sc iv}$\lambda$4658 would be 0.316$\pm$0.008, indicating that this stellar C emission line is blended with nebular [Fe\,{\sc iii}]$\lambda$4658 emission line. Also, the O\,{\sc ii}\,$\lambda$4649 emission line (Table\,4) includes a contribution from the stellar C\,{\sc iii}$\lambda$4650 recombination line with and intensity of 0.385$\pm$0.026 (see Appendix\,\ref{sec:orl_physical}). The presence of these emission lines suggests that the central star of Vy\,2-2 is a weak emission line star ({\it wels}) while the C emission lines indicate a C-rich central star, which strongly contrasts with the O-rich nature of the nebula (Sec.\,\ref{sec:physical}). Nevertheless, other possibilities will be discussed below (Sec.\,\ref{sec:binary}).

\begin{table}
	\caption{Intrinsic intensities of the identified stellar emission lines or blends in units of $I$(H$\beta$)=100.} 
    % title of Table
	\label{table:wels}      % is used to refer this table in the text
	\centering                          % used for centering table
	\begin{tabular}{l c c }        % centered columns (4 columns)
		\hline    
		\rule{0pt}{2.4ex}             % inserts double horizontal lines
		Emission line\tablefootmark{a}  & $f(\lambda)$  & $I(\lambda)$ \\
		\hline                        % inserts single horizontal line
		\rule{0pt}{2ex}
		N\,{\sc iii} (*) $\lambda$4631 & 0.053 & 0.271$\pm$0.018 \\
		\rule{0pt}{2ex}
		O\,{\sc ii} + N\,{\sc iii} (*) $\lambda$4640 & 0.052 & 0.718$\pm$0.019 \\
		\rule{0pt}{2ex}
		O\,{\sc ii} + C\,{\sc iii} (*) $\lambda$4649  & 0.052 & 0.985$\pm$0.019 \\
		\rule{0pt}{2ex}
		[Fe\,{\sc iii}]  + C\,{\sc iv} (*) $\lambda$4658 & 0.050 & 1.404$\pm$0.026 \\
		\rule{0pt}{2ex}
		He\,{\sc ii} (*) $\lambda$4686 & 0.046 & 1.736$\pm$0.031  \\
		\rule{0pt}{2ex}		
		C\,{\sc iv} (*) $\lambda$5801 & -0.201 & 0.253$\pm$0.006 \\
		\rule{0pt}{2ex}
		C\,{\sc iv} (*) $\lambda$5812 & -0.203 & 0.231$\pm$0.011 \\
		
		\hline                                   %inserts single line
	\end{tabular}
\tablefoot{
  \tablefootmark{a}{Stellar emission lines are marked with (*), other emission lines are nebular.}
  }

\end{table}

\subsection{Internal kinematics of the nebula}
\label{sec:kinematics}

\begin{figure*}
	\centering
	\includegraphics[width=5.7cm]{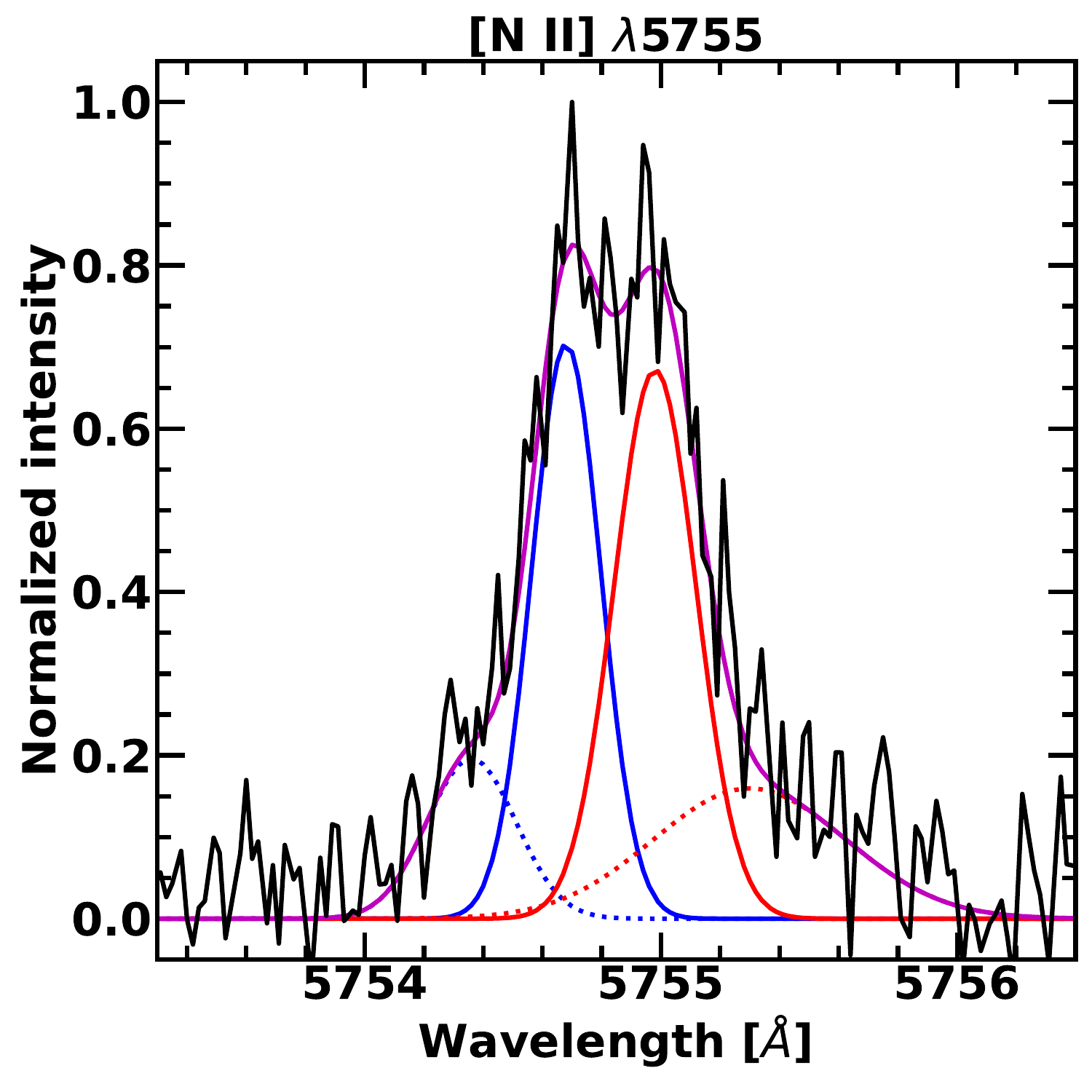}
	\includegraphics[width=5.7cm]{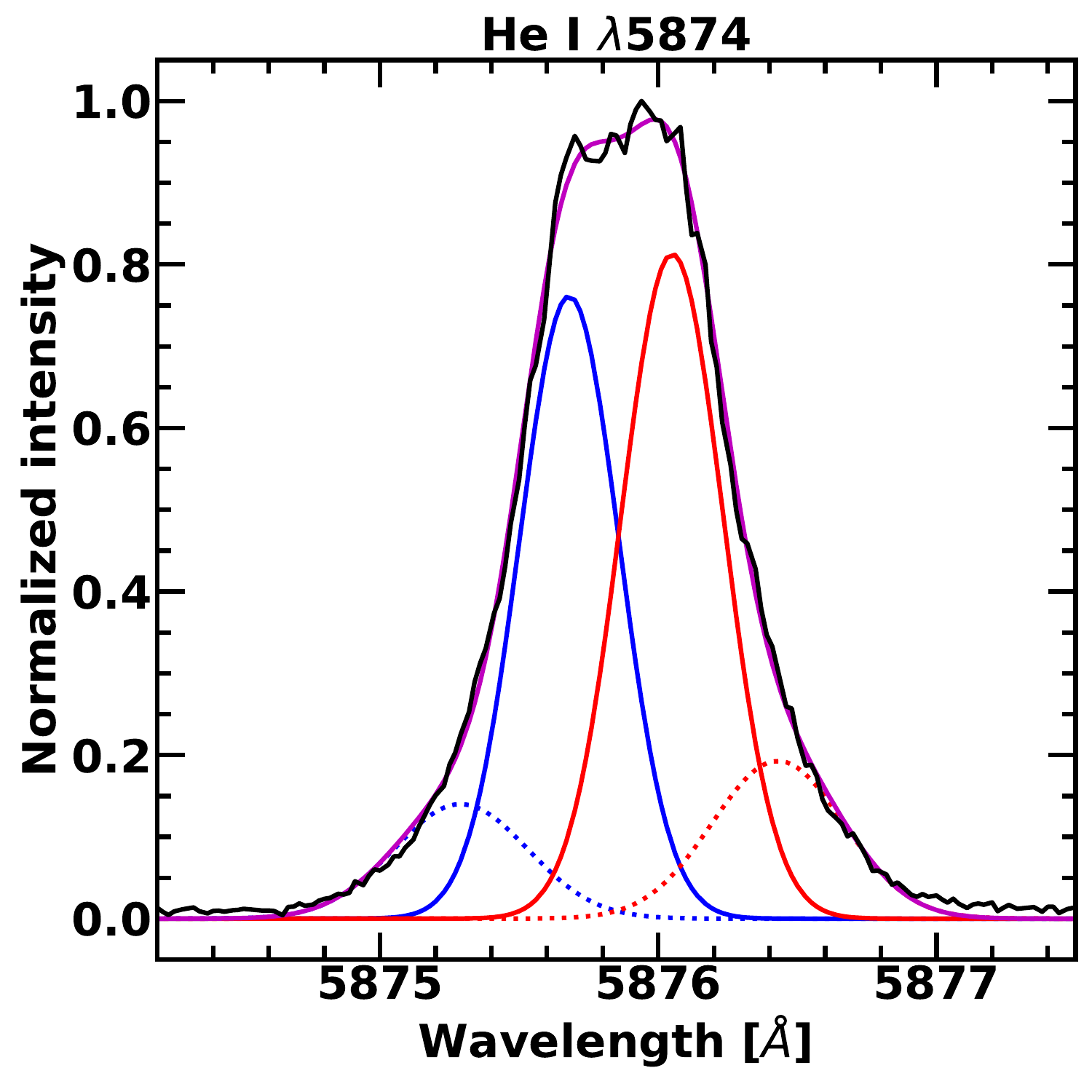}
        \includegraphics[width=5.7cm]{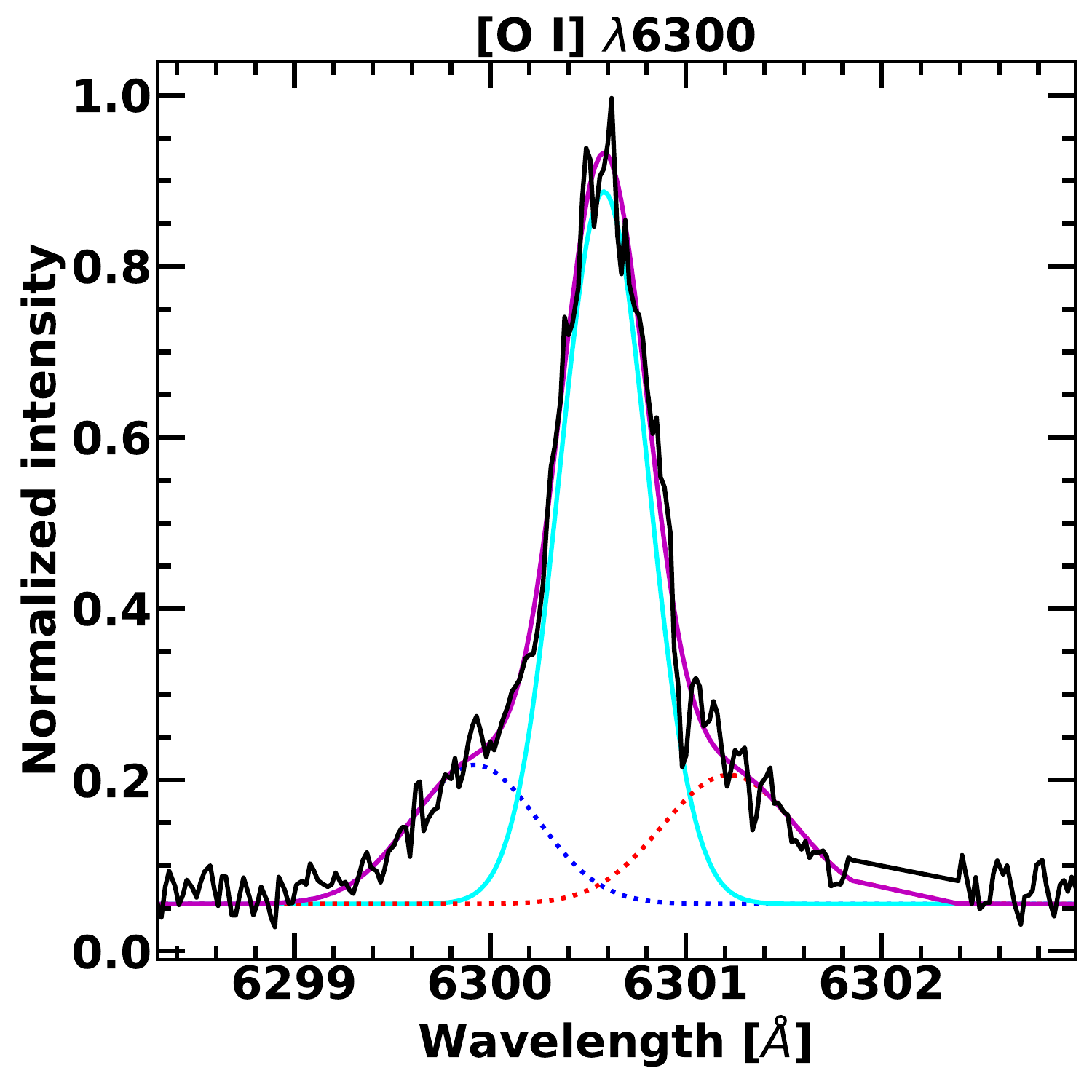}
	\includegraphics[width=5.7cm]{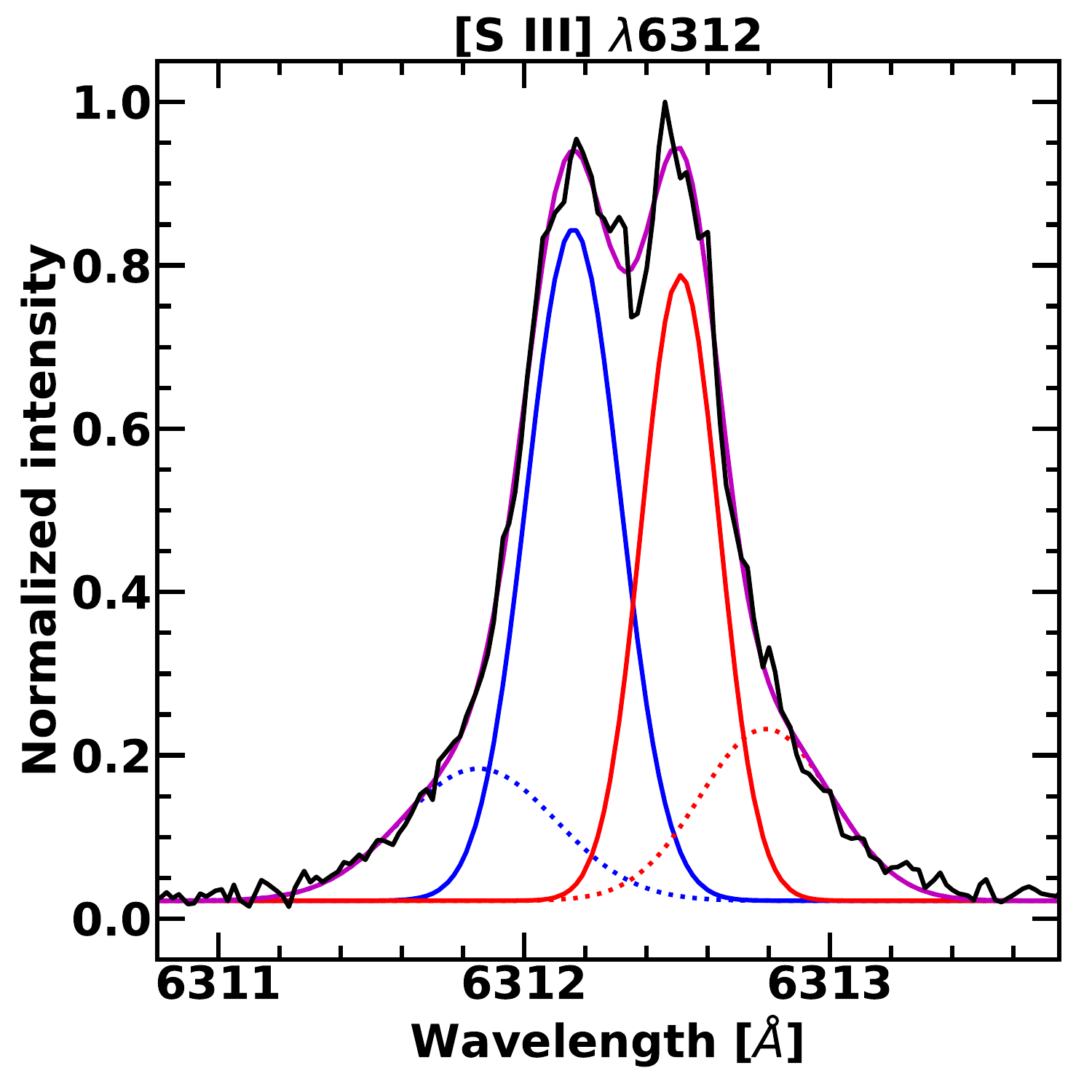}
	\includegraphics[width=5.7cm]{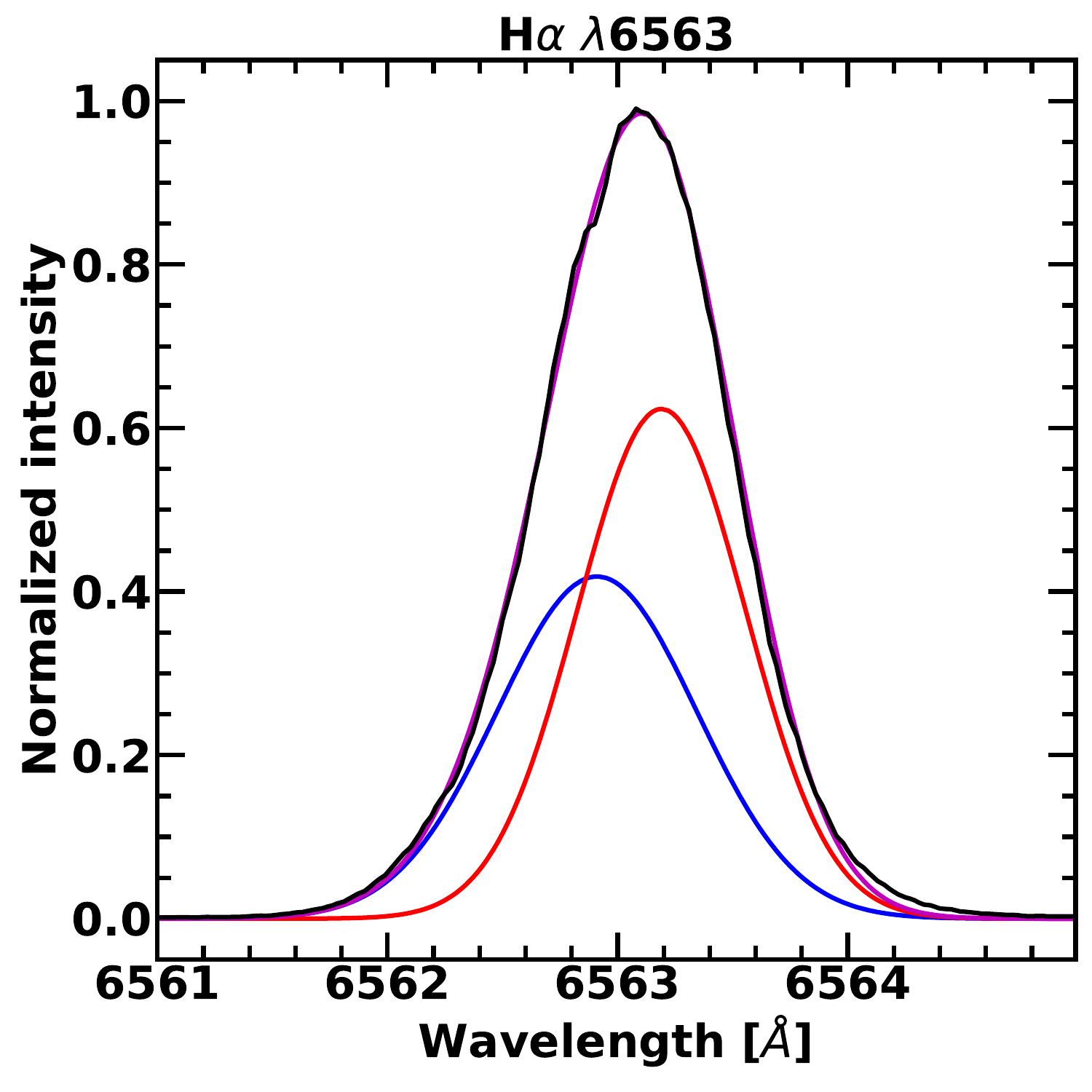}
	\includegraphics[width=5.7cm]{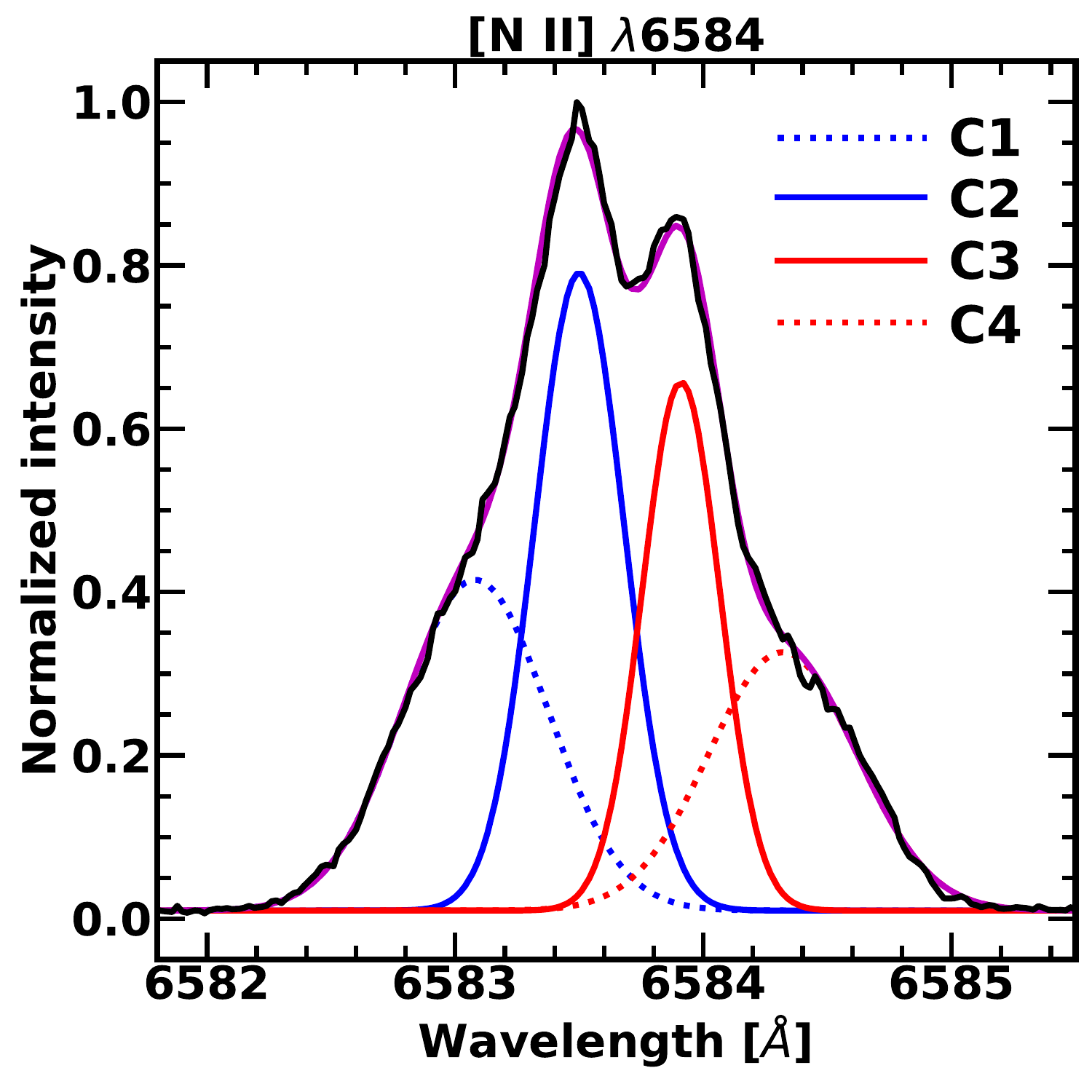}
	\includegraphics[width=5.7cm]{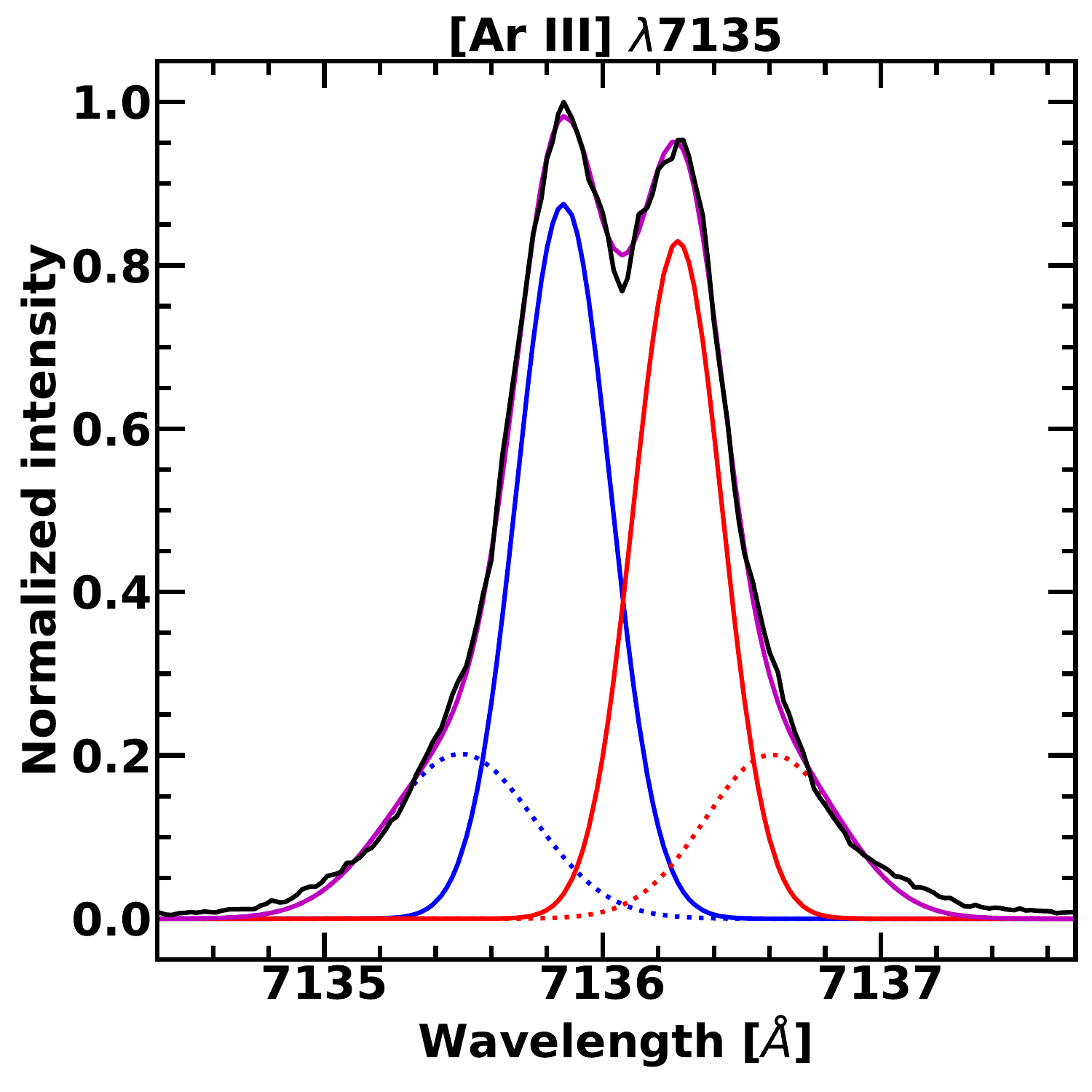}
	\includegraphics[width=5.7cm]{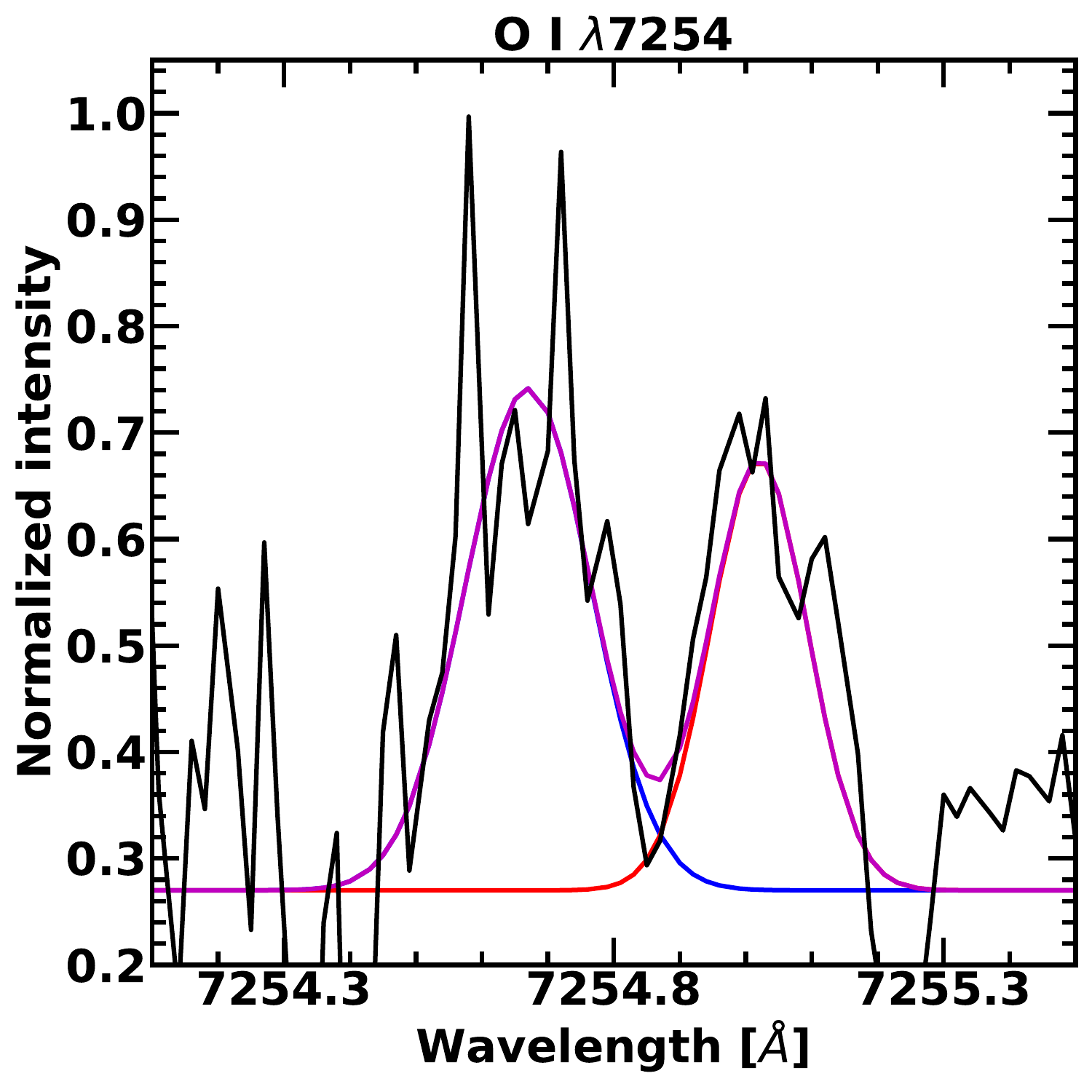}
	\includegraphics[width=5.7cm]{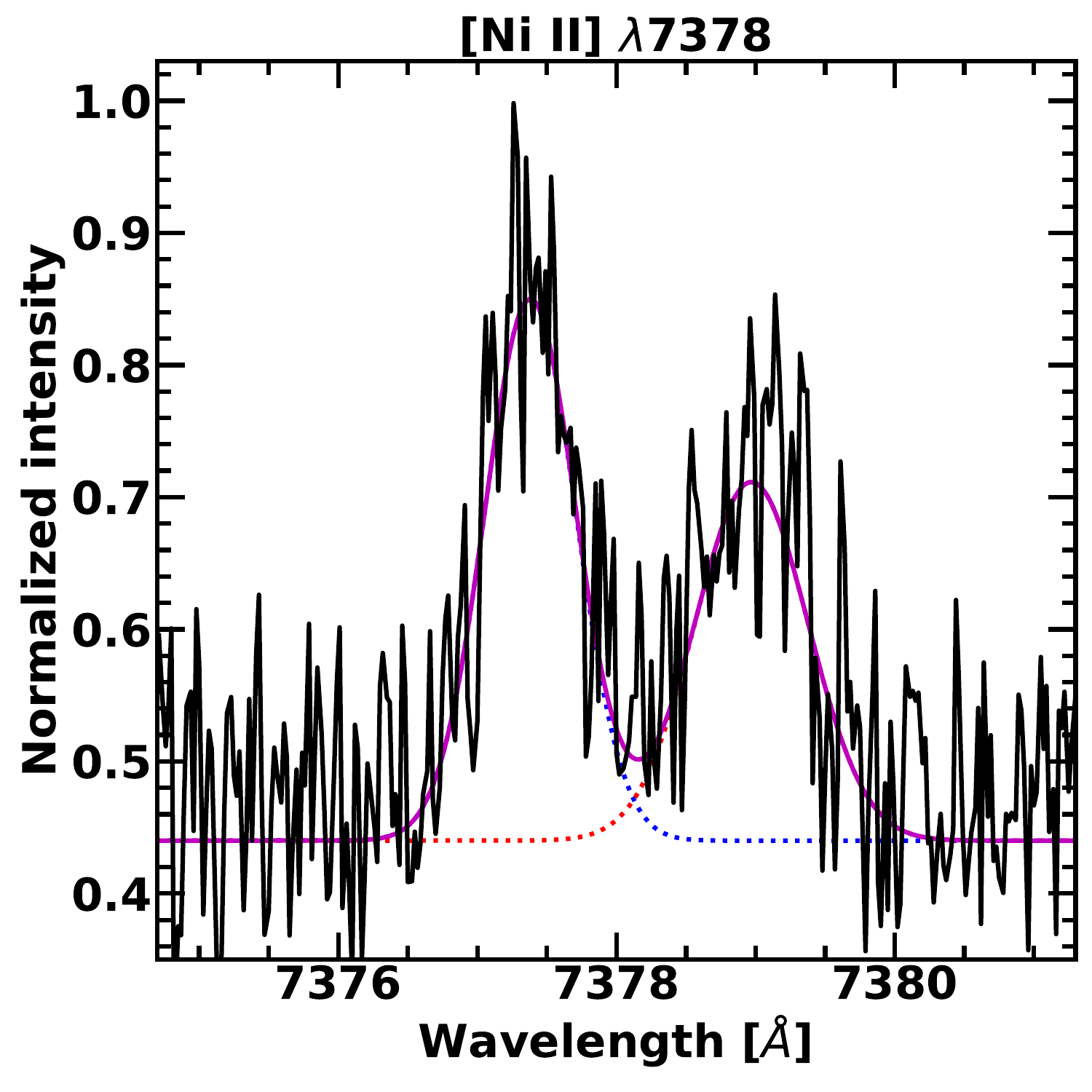}
	\includegraphics[width=5.7cm]{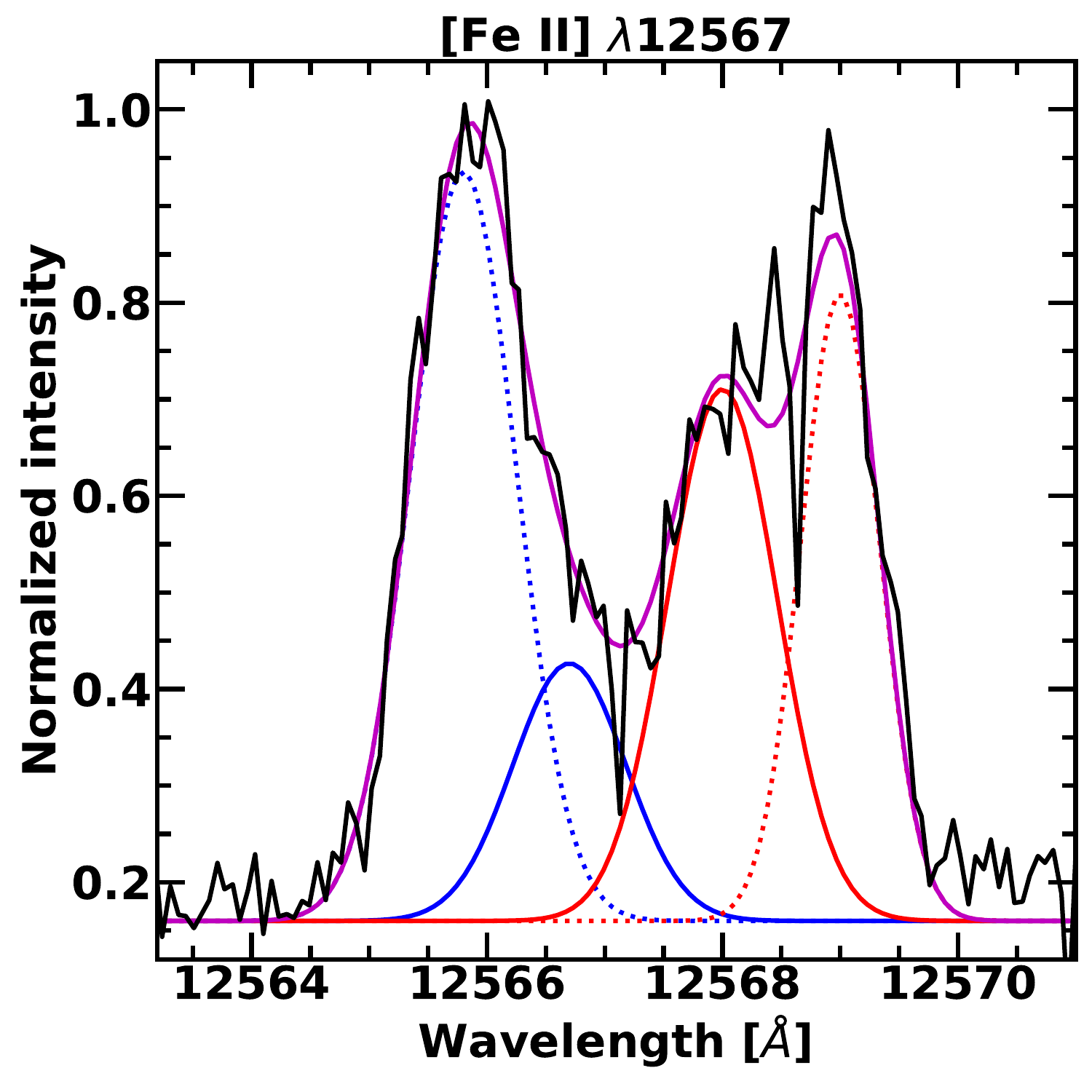}
        \includegraphics[width=5.7cm]{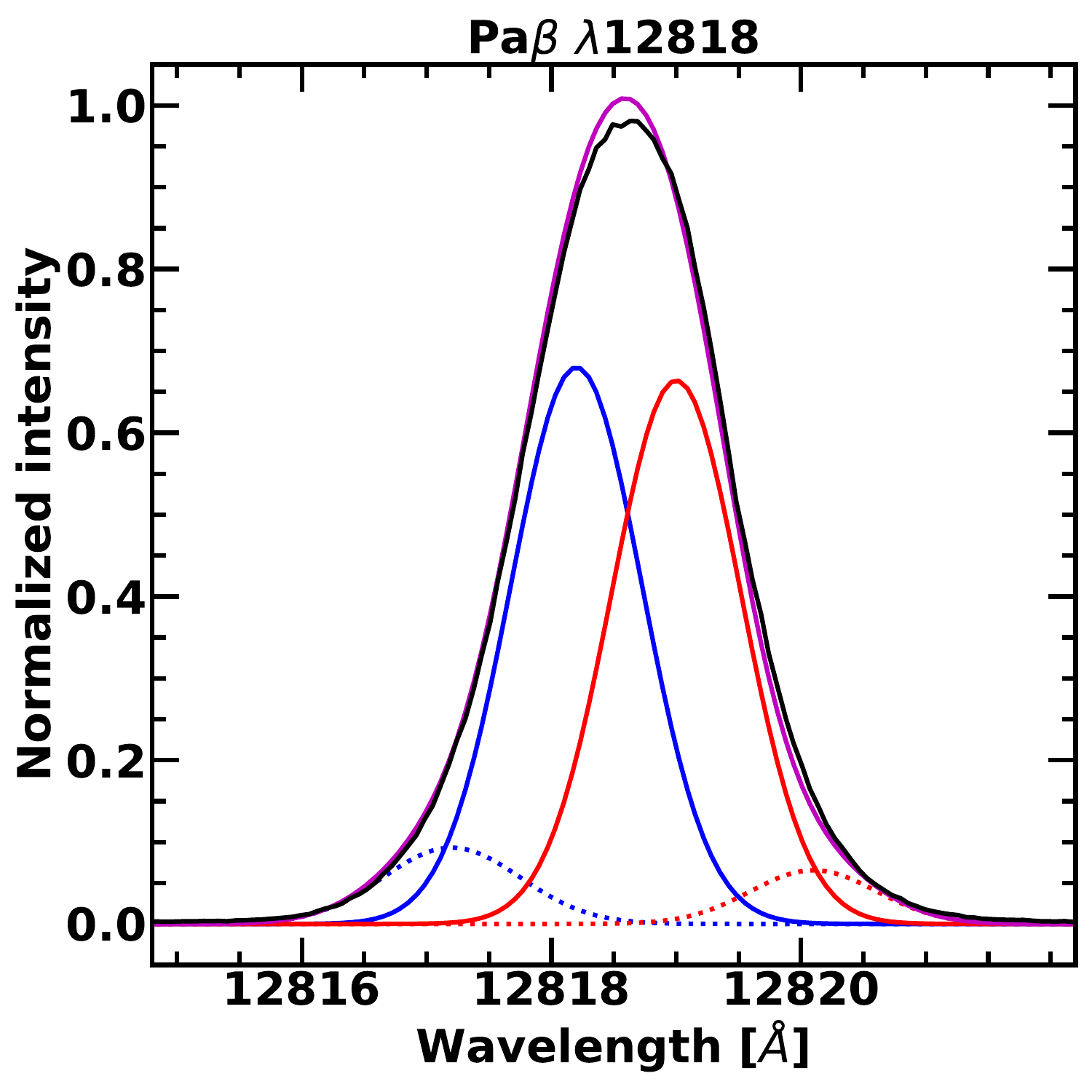}
	\includegraphics[width=5.7cm]{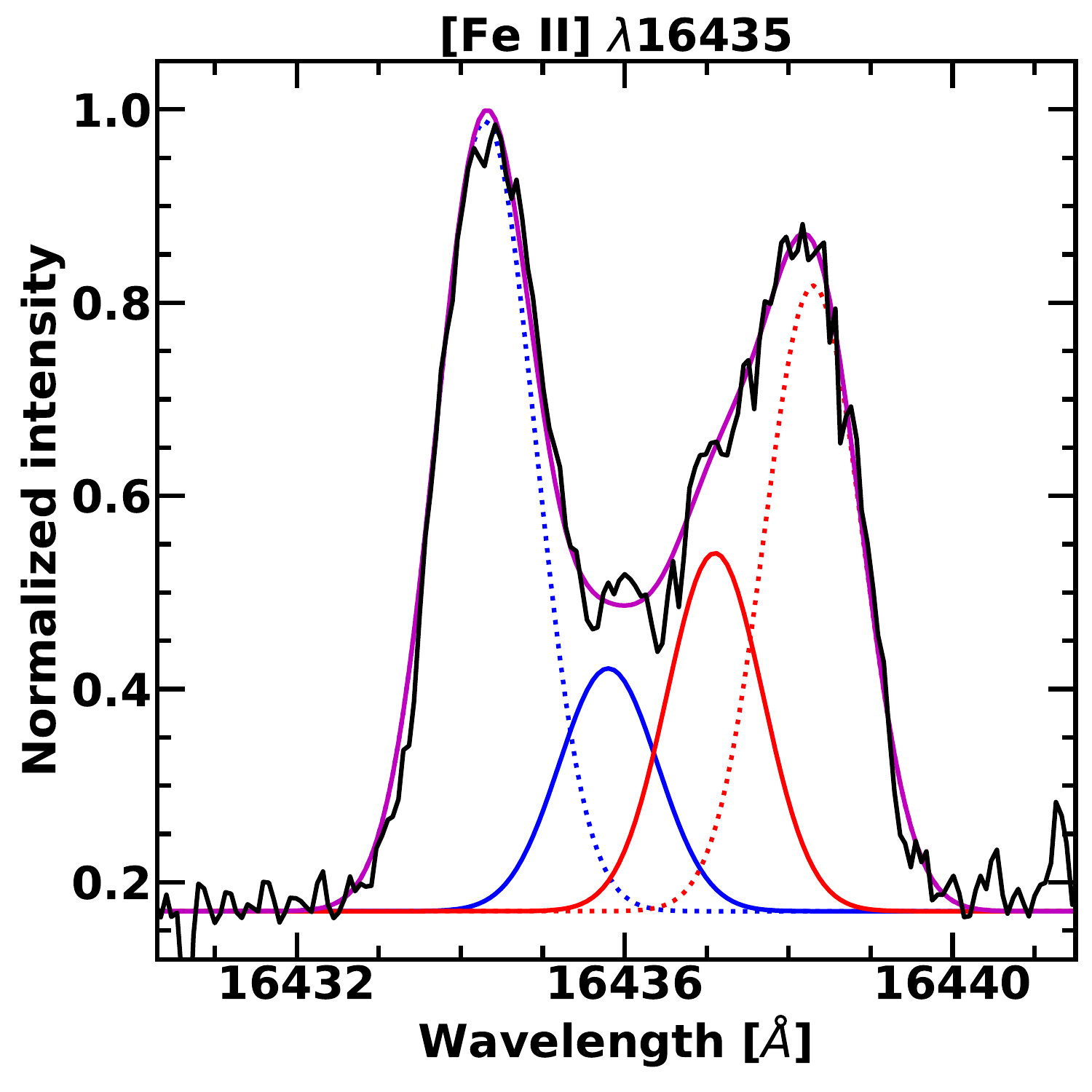}
	\caption{Profiles of selected emission lines in the optical and near-infrared high-resolution CARMENES spectra of Vy\,2-2. Each emission line is identified at the top of its individual panel. The black line is the observed spectrum. The solid and dotted lines represent the central region (components C2--C3) and bipolar lobes (C1--C4), respectively (see the panel of the [N\,{\sc ii}]$\lambda$6585 emission line). Red and blue colors are used for blueshifted and redshifted emission, respectively, with respect to the systemic velocity of Vy\,2-2 ($\upsilon_{\rm sys}(LSR)$ = $-$50.7\,km\,s$^{-1}$).  In the [O\,{\sc i}] emission line, the cyan line represents the blended C2 and C3 components.} 
	\label{fig:CARMENES}
\end{figure*}

In Figure \ref{fig:CARMENES} we show a representative set of emission line profiles observed in the final CARMENES spectrum, including recombination  
and forbidden emission lines of different ionization potential and state. The high spectral resolution achieved by CARMENES (Sec.\,\ref{sec:carmen}) allows us to identify asymmetries and emission peaks in the line profiles, which are indicative of the existence of different kinematical components in the nebula. In particular, two or four different kinematical components can be recognized, depending on the emission line.

To analyze and characterize the kinematical components we have performed a multi-gaussian line fit to the emission line profiles assuming two or four gausssians. We will use the wavelength at the intensity peak and the FWHM of each gaussian as representative of the radial velocity [$\upsilon$$_{\rm r}$(LSR), where LSR stands for Local Standard of Rest] and the velocity width [$\Delta$$\upsilon$] of each component, respectively. The results are listed in Table \ref{table:CARMENES}. An inspection of Table \ref{table:CARMENES} reveals that the radial velocity defines four different kinematical components, hereafter refereed to as C1, C2, C3, and C4,  with radial velocities around $-80$, $-60$, $-40$, and $-20$\,km\,s$^{-1}$, respectively. These components are also indicated in Table \ref{table:CARMENES}. Moreover, the pairs C1--C4 and C2--C3 appear symmetric with respect to the radial velocity centroid of the emission profiles. The velocity width of C1 and C4 typically ranges from $\sim$15 to $\sim$30\,km\,s$^{-1}$, which are values smaller than for C2 and C3. Components C1 to C3 were already identified by \cite{mir91} in the [N\,{\sc ii}]$\lambda$6584 emission line profile, but they did not identify C4 probably due to the lower spectral resolution in their spectra (FWHM$\simeq$6\,km\,s$^{-1}$) than in four CARMENES ones (Sec.\,\ref{sec:carmen}). 

Only C2 and C3 could be unambiguously identified in the H$\alpha$ emission line. This is most probably because H$\alpha$ shows broad emission wings that extend up to $\sim$1500\,km\,s$^{-1}$ (not show here) and hid the weaker C1 and C4 components. Extended H$\alpha$ wings in Vy\,2-2 were already reported by \cite{arr03} who found a velocity extent similar to ours and attributed them to Rayleigh-Raman scattering. Conversely, the Paschen lines are not affected by this scattering and the four components are recognized. We have not included the [S\,{\sc ii}]$\lambda$6716,6731 lines (not shown here) in our analysis because they are extremely weak and have a very low signal-to-noise ratio. The [O\,{\sc i}]$\lambda$6300,6363 lines present a peculiar profile (Fig.\,4): C1 and C4 are clearly present but a single emission peak is observed where the two central emission peaks due to C2 and C3 would be expected. We tried to fit these lines using two Gaussian profiles at their center, but no reasonable result was obtained. In Table \ref{table:CARMENES} we assign the measurements to C2-C3. In the O\,{\sc i}$\lambda$7256 line only C2 and C3 can be identified while in the [Ni\,{\sc ii}]$\lambda$7380 line only C1 and C4 can be recognized.  

The intensity of the He\,{\sc i} and H\,{\sc i} emission lines is largely dominated by C2--C3, and a similar situation is observed in the relatively high excitation emission lines of [Ar\,{\sc iii}] and [S\,{\sc iii}] (Fig.\,4). In [N\,{\sc ii}] the intensity difference between C2--C3 and C1--C4 is not so marked, while C1–C4 clearly dominate the low-excitation lines of [Ni\,{\sc ii}] and [Fe\,{\sc ii}]. It is evident that C2--C3 are related to regions with higher excitation than those where C1--C4 arise. 

Finally, in order to obtain the systemic velocity ($\upsilon_{\rm sys}$) of Vy\,2-2, we firstly compute the mean value of the radial velocities of the components identified in all emission lines. Afterwards, we  determined two different values for $\upsilon_{\rm sys}$: the first one using C1 and C4, and the second one using C2 and C3. These three values of $\upsilon_{\rm sys}$ were found to be virtually identical to each other, indicating a noticeable symmetry in the internal kinematics of Vy2-2. We finally report $\upsilon_{\rm sys}$(LSR)\,=\,$-50.7$$\pm$0.1\,km s$^{-1}$.

\begin{figure}
	\centering
	\includegraphics[width=7.4cm]{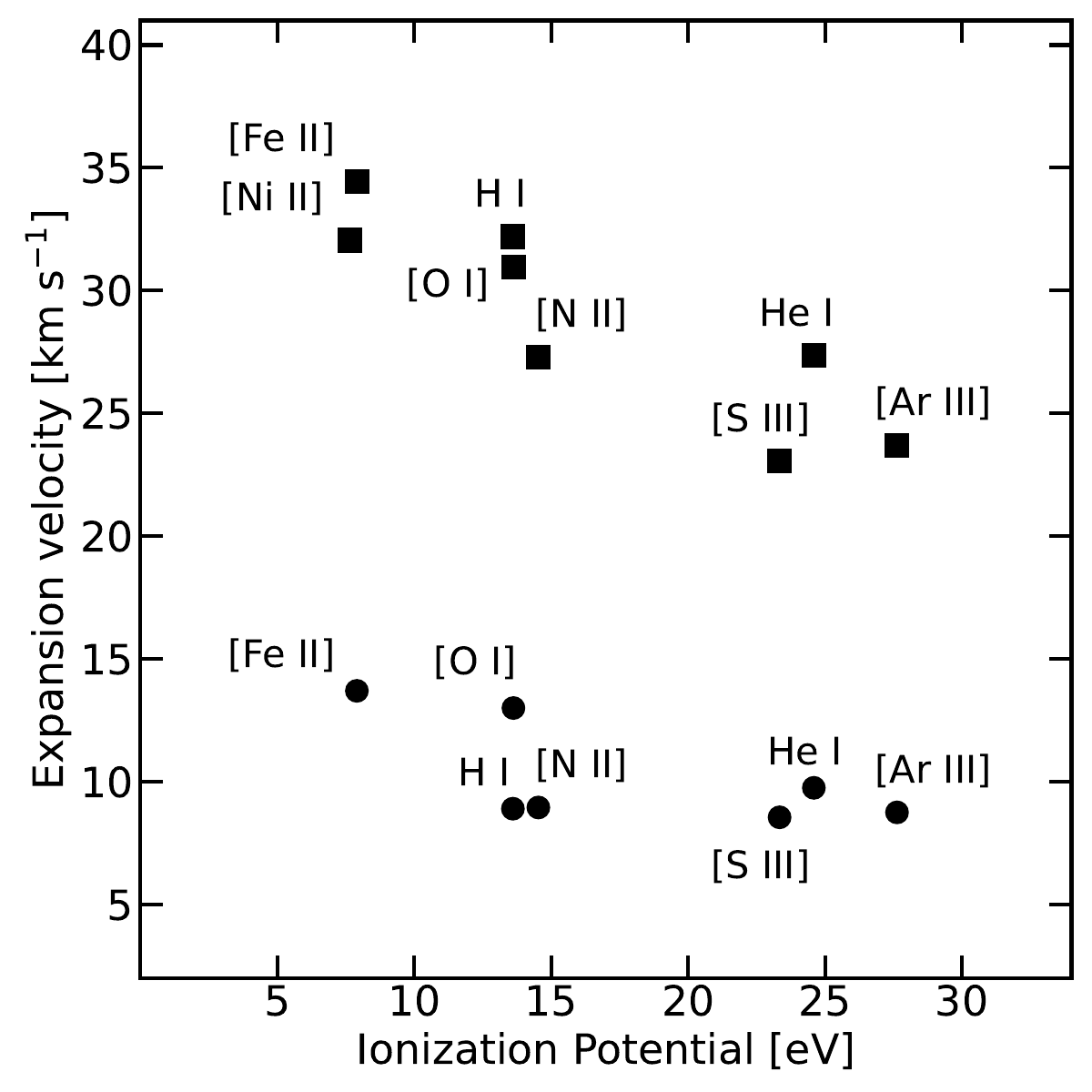}
	\caption{Expansion velocity versus ionization potential in the bipolar lobes (filled squares) and bright central region (filled circles) of Vy 2-2 (see Table\,\ref{table:exp_vel}).}
	\label{fig:wilson}
\end{figure}

\begin{table}
	\caption{Mean ionic expansion velocities ($\upsilon$$_{\rm e}$) of each spectral component of the ionized gas shown in Table\,\ref{table:CARMENES}, with respect to the systemic velocity $\upsilon_{\rm sys}$ = $-50.7$ km s$^{-1}$.} % title of Table
	\label{table:exp_vel}      % is used to refer this table in the text
	\centering                          % used for centering table
	\begin{tabular}{c c c}        % centered columns (4 columns)
		\hline
		\rule{0pt}{2.5ex}                 % inserts double horizontal lines
		Ion & Component  & $\upsilon$$_{\rm e}$ [km s$^{-1}$]\\    % table heading 
		\hline                        % inserts single horizontal line
		\rule{0pt}{3ex}
		H$^{+}$   & 1 & $-$31.5   \\
		& 2 & $-$8.8   \\
		& 3 & $+$9.0   \\
		& 4 & $+$32.9   \\
		\rule{0pt}{2ex}
		He$^{+}$  & 1 & $-$29.3   \\
		& 2 & $-$10.1   \\
		& 3 & $+$9.4   \\
		& 4 & $+$25.4   \\
		\rule{0pt}{2ex}
		N$^{+}$   & 1 & $-$28.0   \\
		& 2 & $-$9.5   \\
		& 3 & $+$8.4   \\
		& 4 & $+$26.6   \\
		\rule{0pt}{2ex}
		O$^{0}$ & 1 & $-$31.0   \\  
            & 2 & $-$13.0\tablefootmark{a}   \\
            & 3 & $+$13.0\tablefootmark{a}   \\
		& 4 & $+$30.9   \\
		\rule{0pt}{2ex}
%		S$^{+}$   & 2 & ($-$30.1)   \\
%		& 3 & ($+$13.4)   \\
		\rule{0pt}{2ex}
		S$^{2+}$  & 1 & $-$22.9   \\
		& 2 & $-$8.2   \\
		& 3 & $+$8.4   \\
		& 4 & $+$21.7   \\
		\rule{0pt}{2ex}
		Ar$^{2+}$  & 1 & $-$25.7   \\
		& 2 & $-$9.8   \\
		& 3 & $+$7.7   \\
		& 4 & $+$21.7   \\	
		\rule{0pt}{2ex}
		Fe$^{+}$  & 1 & $-$33.5   \\
		& 2 & $-$10.1   \\
		& 3 & $+$17.3   \\
		& 4 & $+$35.4   \\
		\rule{0pt}{2ex}
		Ni$^{+}$  & 1 & $-$30.1   \\
		& 4 & $+$34.0   \\
		
		\hline                                   %inserts single line
	\end{tabular}
\tablefoot{
  \tablefootmark{a}{Obtained from $\Delta$$\upsilon$ of component 2--3 of the [O\,{\sc i}]$\lambda$6300 in Table\,\ref{table:CARMENES} (see text for details).}
  }
\end{table}

\section{Discussion}

\subsection{The morphokinematical structure of Vy\,2-2}
\label{sec:3d_morpho}

MS91 detected three components in the line profile of [N\,{\sc ii}] spectra, and proposed a simple model for the 3D structure of Vy\,2-2, namely, a slowly expanding central region and two wide bipolar lobes. The high resolution of the CARMENES spectra allowed us to detect four components and to refine the model. A comparison of the [N\,{\sc ii}] line profile in Fig.\,\ref{fig:CARMENES} with the position-velocity maps of the [N\,{\sc ii}] emission line in MS91 (e.g., their Fig.\,3) shows that C2--C3 are located at the center of the emission line and should correspond to the central region, C1 is located towards the NE, and C4 towards the SW, consequently, C1--C4 trace the bipolar lobes.  Remarkably, while the existing images of Vy\,2-2 do not provide information about properties of the nebular components, our CARMENES spectra allow us to state that the central region presents a higher excitation than the bipolar lobes, where a very low excitation should exist, as suggested by the presence of [Fe\,{\sc ii}] and [Ni\,{\sc ii}] emission lines with the highest expansion velocities (see also next paragraph). 

Assuming this morphological structure we can obtain the expansion velocity of each kinematic component in each ion. Table\,5 shows the results where we have averaged the values from emission lines of the same ion. To further analyze these results, we have plotted the expansion velocity versus the ionization potential of each ion in the central region and bipolar lobes. The results are presented in Fig.\,\ref{fig:wilson} and show the well-known Wilson effect in both structures, with the expansion velocity increasing for decreasing ionization potential \citep{wil50}. This strongly suggests a stratification in the nebula such that the low/high excitation lines arise in the outer/inner nebular regions. These results are typical of a photoionized nebula. In fact, we have not found any evidence in our optical spectra for shock excitation in the main nebula, which is also indicated by the near-infrared spectra by \cite{kap26} who found shock excitation only in the bipolar jets/bullets. In particular,  we note that the intensities of the [Fe\,{\sc ii}]$\lambda$8761 and [Ni\,{\sc ii}]$\lambda$7378 emission line place Vy\,2-2 in the photoionized region of the $log$([Ni\,{\sc ii}]$\lambda$7378/H$\alpha$) versus $log$([Fe\,{\sc ii}]$\lambda$8761/H$\alpha$) obtained by \citet[their Fig. 15]{bou25}. Moreover, the intensities of the near-infrared Br$\gamma$, He\,{\sc i} 2.058$\mu$m, and He\,{\sc i} 2.120$\mu$m recombination lines obtained by \cite{hora99} in Vy 2-2 place it
in a position incompatible with shock-ionization models, according to the
diagnostic diagram presented by \cite{cal25b}. Corroborating that shock excitation does not dominate or is not relevant in our spectrum of the main ionized shell of Vy\,2-2 (and in any other PN or structure) is a crucial step to confirm whether the classical method to obtain chemical abundances is applicable. If shocks play an important role in the observed spectrum, this classical method may provide unrealistic results \citep{cal24b}.

Assuming a spherical shell of radius $\simeq$0$\farcs$24, as suggested by the 14.965 and 22.485\,GHz radio continuum emission maps, and an expansion velocity of 9\,km\,s$^{-1}$ in H$^+$ for the central region, we obtain a kinematical age of $\simeq$125\,$\times$\,\textit{D}\,[kpc]\,yr where \textit{D} is the distance to the nebula. For Vy\,2-2, \cite{chr98} obtained a value of 1.13$\pm$0.12\,mas\,yr$^{-1}$ for the angular expansion of the central region as observed in radio continuum. Combining this value with the expansion velocity of the central region in H$^+$, and assuming spherical geometry, we derive a distance of $\simeq$1.68$\pm$0.02\,kpc, implying a kinematical age of 210$\pm$3\,yr, which further corroborates the general idea that Vy\,2-2 is an extremely young PN.  

\subsection{The formation of Vy\,2-2 and its central star }      
\label{sec:binary}

The central star of Vy\,2-2 had never been directly identified before and only some values for its possible effective temperature (41000--59000\,K) were inferred from the nebular emission lines \citep{kal91}. The stellar emission lines clearly identified in our spectra would be the first direct detection of this central star and its {\it wels} spectral type. 

However, \citet{par98} and \citet{wei15} have already shown that \textit{wels} may not be of real spectral type of central stars because absorption lines are detected in some {\it wels} when observed at enough spectral resolution, indicating different classifications. Actually, observations of several PNe have shown that stellar emission lines, as those observed in Vy\,2-2 (Fig.\,3), arise from main-sequence companions irradiated by the central stars in post-common envelope binary systems \citep[e.g.,][]{mis11a, mis11b, jon17, mun20}. Therefore, the possibility is open that Vy\,2-2 hosts a post-common envelope binary.

Interestingly, other properties of Vy\,2-2 also point to a close binary. An unusually high $adf$(O$^{2+}$), as derived in Vy\,2-2 (Sect.\ref{sec:physical}), has also been found by \cite{wes18} in close binary systems (period around 1\,day) with values of $adf$(O$^{2+}$)$\geq$10. Circular concentric around the main nebula, bipolar morphologies, and collimated outflows are signatures of binary stars \citep{iva13, zou20, sok25} and these structures are observed in Vy\,2-2 \citep[Fig.\,1 and Sect.\,4.1]{ram16, kap26}. Fluorine production may also be related to binary star evolution \citep{ots08}. Finally, the Rayleigh$-$Raman scattering traced by the large H$\alpha$ wings requires the existence of a very high density ($\geq$10$^9$\,cm$^{-3}$) region located very close to the central star, which could probably be better explained by invoking a companion \citep[][and references therein]{mir22}. 

If the observed stellar emission lines are associated to a companion, the central star of this PN remains unidentified. Fortunately, estimates for its $M_{\rm i}$ may be obtained from observed abundance ratios and other observational data. The value of C/O=0.15$^{+0.09} _{-0.04}$ in Vy\,2-2 (Sec.\,\ref{sec:physical}) is extremely low and, to the best of our knowledge, the lowest one ever found in PNe. According to \cite{mil16} (see his Table\,2), only stellar models with $M_{\rm i}$\,$\lesssim$\,0.85\,M$_{\odot}$ and metallicities $Z$=1-10$\times$10$^{-3}$ will end up in the AGB phase with such a low value of C/O surface abundance ratio. The subsolar metallicity found in  Vy\,2-2 is compatible with those stellar models while the low $M_{\rm i}$ is compatible with the O-rich nature of the nebula and very low N/O abundance ratio. Nevertheless, we note that the C/O abundance ratio of O-rich AGB stars increases after each thermal pulse and that a low C/O abundance ratio could also be caused if the AGB evolution of the progenitor of the central star is interrupted due to interaction with a companion \citep[e.g.][]{deM09, kar14, mil16}. 

The $M_{\rm i}$ of the central star can also be estimated by calculating the mass of the main nebular components, namely, the ionized gas, the molecular gas, the photodissociation region (PDR), and the dust. Table\,\ref{table:mass} lists the estimated mass of these components by using already reported values and/or available data. We note that the PDR is the heaviest component, followed by the molecular and ionized gas, while dust is the lightest one. With the distance to Vy\,2-2 determined in Section\,\ref{sec:3d_morpho}, the total mass of the circumstellar nebula ($M_{\rm CSE}$) is 4.79$\pm$0.11\,$\times$\,10$^{-2}$\,M$_{\odot}$. In addition, if we assume that the mass lost by the star during the red giant branch phase ($M_{\rm RGB}$) is 0.2$\pm$0.1\,M$_{\odot}$ \citep{dor93}, the mass lost during the previous evolutionary phases ($M_{\rm lost}$=$M_{\rm CSE}$+$M_{\rm RGB}$) would be 0.25$\pm$0.10\,M$_{\odot}$. Using, as a first approximation, evolutionary models for single stars \citep{mil16}, we obtain an initial mass for the progenitor star of Vy\,2-2 of $\lesssim$0.9\,M$_{\odot}$, implying a central star mass of $\simeq$0.5--0.6\,M$_\odot$ \citep{mil16}. If the stellar emission lines detected in Vy\,2-2 indeed arise from an irradiated companion with a mass of 0.1--0.6\,M$_\odot$, as found in other $\it wels$ binaries \citep{mis11a, mis11b, jon17, wes18}, orbital periods of 0.3--3.0\,d and orbital separations of $\simeq$1.5--9.5\,R$_{\odot}$ could be expected \citep{wes18}. Time-resolved spectroscopic observations of the stellar emission lines to look for periodic radial velocity modulations are crucial to confirm the binary nature of the central star of Vy\,2-2 and to constrain some of its properties. 

\begin{table}
	\caption{Estimated mass of the main nebular components in Vy\,2-2 and mass ratios between some of them.} % title of Table
	\label{table:mass}      % is used to refer this table in the text
	\centering                          % used for centering table
	\begin{tabular}{l r}        % centered columns (4 columns)
		\hline    
		%\rule{0pt}{2.4ex}             % inserts double horizontal lines
			 Mass$_{component}$ & Mass [M$_{\odot}$] \\    % table heading 
		\hline                        % inserts single horizontal line
		%\rule{0pt}{2.5ex}
		$M_{\rm ion}$\tablefootmark{a} &  1.6$\times$10$^{-4}$$\times$$D$$^{2.5}$ \\
		%\rule{0pt}{2.5ex}
		$M_{\rm PDR}$\tablefootmark{b} &  1.6$\times$10$^{-2}$$\times$$D$ $^{2}$ \\
		%\rule{0pt}{2.5ex}
		$M_{\rm m}$\tablefootmark{c}  & 7.0$\times$10$^{-4}$$\times$$D$ $^{2}$ \\
		%\rule{0pt}{2.5ex}
		$M_{\rm dust}$\tablefootmark{a} & 1.4$\times$10$^{-6}$$\times$$D$ $^{2}$ \\
				
		\hline
				
		\hline
		%\rule{0pt}{2.5ex}
        $M_{\rm CSE}$\tablefootmark{d} &  4.8$\times$10$^{-2}$ \\
		%\rule{0pt}{2.5ex}
        $M_{\rm RGB}$\tablefootmark{e} &  0.2 \\
		\hline
		%\rule{0pt}{2.5ex}
		$M_{\rm lost}$\tablefootmark{f} &  0.25 \\
		\hline   
        %inserts single line
	\end{tabular}
     \tablefoot{
 \tablefootmark{a}{Ionized ($M_{\rm ion}$) and dust ($M_{\rm dust}$) mass reported by \cite{dav79}.}
  \tablefootmark{b}{Mass of the PDR ($M_{\rm PDR}$) obtained from the [C\,{\sc ii}]$\lambda$158\,$\mu$m flux \citep{liu01}, C/H=4.661$\times$10$^{-4}$ \citep{wes05}, and the expression in \cite{liu01}.}
  \tablefootmark{c}{Molecular mass ($M_{\rm m}$) determined by \cite{hug96}.}
  \tablefootmark{d}{Total circumstellar mass ($M_{\rm CSE}$) as the sum of the previous values, adopting $D=1.68$$\pm$0.02\,kpc (see Sec.\,\ref{sec:3d_morpho}).}
  \tablefootmark{e}{Typical mass lost in the red giant branch phase ($M_{\rm RGB}$) \citep{dor93}.}
  \tablefootmark{f}{Total mass lost $M_{\rm lost}$=$M_{\rm CSE}$+$M_{\rm RGB}$.}
  }
\end{table}

\section{Conclusions}

We have analyzed new intermediate- and \'echelle high-resolution spectra of the OHPN Vy\,2-2 with the idea of obtaining a new value for the \textit{adf} for this object, investigating the internal kinematics and structure of the nebula, and inferring its possible formation processes. The main conclusions of this work can be summarized as follows.

Our intermediate-resolution spectra have revealed many previously unreported nebular recombination (e.g., N$^{+}$) and forbidden (e.g., [Ni\,{\sc ii}] and, possibly, [F\,{\sc iv}]) emission lines, as well as stellar recombination emission lines of C$^{3+}$, C$^{4+}$, and N$^{3+}$. Some of the newly detected nebular recombination emission lines are crucial in determining abundances from ORLs.

The analysis of the intermediate-resolution spectra corroborates that Vy\,2-2 is composed by several regions characterized by their own $T_{\rm e}$ and $N_{\rm e}$, with an inner region where the electron density is very high ($\simeq$65000\,cm$^{-3}$). Additionally, the broad Rayleigh–Raman scattering H$\alpha$ wings observed in the high-resolution spectrum strongly suggest the existence of an extremely dense region ($\geq$10$^9$\,cm$^{-3}$) in the immediate vicinity of the central star. 

The abundances determined from CELs are subsolar and similar to those obtained previously by other authors. The new value of the $adf$(O$^{2+}$) obtained from our intermediate-resolution spectra is 13.6$\pm$2.6 which favors including Vy\,2-2 among the PNe with unusually high $adf$(O$^{2+}$).

The high-resolution spectra allow us to analyze the emission lines with different ionization degree and ionization potential. We identify four kinematical components in the nebula: two components related to the relatively high-excitation central region and two components tracing the relatively
low-excitation bipolar lobes of the object. The kinematics observed in emission lines with different ionization potential, our optical and published near-infrared spectra indicate that shock excitation is not present in the main ionized shell of Vy2-2. This provides support to apply to Vy2-2 the usual method for calculating chemical abundances.

By combining the expansion velocity of the central region in H$^+$ with the angular expansion of the same region obtained from radio continuum observations, we obtain a distance of $1.68\pm0.02$\,kpc  to Vy\,2-2, which implies a kinematical age of 210$\pm$3yr, further supporting the extreme youth of this PN. 

Although the stellar emission lines detected in Vy\,2-2 suggest a {\it wels} type central star, they may be related to an irradiated close companion. This possibility is favored by the unusually high \textit{adf}, the concentric rings, bipolar morphology, collimated outflows, the origin of the Rayleigh–Raman scattering H$\alpha$ wings, and presence of fluorine. In fact, all these properties suggest that Vy\,2-2 may host a post-common-envelope binary. 

Finally, abundance ratios and an estimate for the mass lost by the star progenitor of Vy\,2-2 during its life suggest a relatively low main-sequence initial mass of $\lessapprox$0.9\,M$_{\odot}$. If the stellar emission lines indeed arise in an irradiated companion, a comparison with similar systems suggests a secondary mass of 0.1--0.6\,M$_\odot$, resulting in a binary system with an orbital period of $\simeq$0.3--3.0\,d and an orbital separation of $\simeq$1.5--9.5\,R$_{\odot}$.

\begin{acknowledgements}
      
      We are very grateful to our anonymous referee for their constructive comments that have improved the paper. CARMENES is an instrument for the Centro Astronómico Hispano-Alemán de Calar Alto (CAHA, Almería, Spain). CARMENES was funded by the Max-Planck-Gesellschaft (MPG), the Consejo Superior de Investigaciones Científicas (CSIC), the Ministerio de Economía y Competitividad (MINECO) and the European Regional Development Fund (ERDF) through projects FICTS-2011-02, ICTS-2017-07-CAHA-4, and CAHA16-CE-3978, and the members of the CARMENES Consortium (Max-Planck-Institut für Astronomie, Instituto de Astrofísica de Andalucía, Landessternwarte Königstuhl, Institut de Ciències de l’Espai, Institut für Astrophysik Göttingen, Universidad Complutense de Madrid, Thüringer Landessternwarte Tautenburg, Instituto de Astrofísica de Canarias, Hamburger Sternwarte, Centro de Astrobiología and Centro Astronómico Hispano-Alemán), with additional contributions by the MINECO, the Deutsche Forschungsgemeinschaft (DFG) through the Major Research Instrumentation Programme and Research Unit FOR2544 “Blue Planets around Red Stars”, the Klaus Tschira Stiftung, the states of Baden-Württemberg and Niedersachsen, and by the Junta de Andalucía. This research is based on observations collected at the Centro Astron{\'o}mico Hispano en Andaluc{\'i}a (CAHA) at Calar Alto, operated jointly by Junta de Andaluc{\'i}a and Consejo Superior de Investigaciones Cient{\'i}ficas (IAA-CSIC). Also, it is based on observations made with the NASA/ESA Hubble Space Telescope obtained from the Space Telescope Science Institute, which is operated by the Association of Universities for Research in Astronomy, Inc., under NASA contract NAS 5-26555. These observations are associated with program 6353. RAC, LFM, JFG are supported by grant PID2023-146295NB-I00, funded by MCIN/AEI/10.13039/501100011033 and FEDER, EU, and by grant CEX2021-001131-S, funded by MCIN/AEI/10.13039/501100011033. RC also acknowledges support by the predoctoral grant PRE2018-085518, funded by MCIN/AEI/10.13039/501100011033 and by ESF Investing in your Future. We thank Pedro J. Amado for helping with the data reduction and data handling. 
      
  \end{acknowledgements}

\begin{appendix}

\section{Physical conditions from optical recombination lines}
\label{sec:orl_physical}

N\,{\sc ii} recombination lines help determine physical conditions using the $\lambda$5679.56/$\lambda$5666.63 and $\lambda5679.56$/$\lambda$4041.31 line intensity ratios \citep{fan11, fan13}. In our spectra, the N\,{\sc ii}\,$\lambda5679.56$ emission line is blended with the N\,{\sc ii}\,$\lambda5676.02$ one, and theoretically $I(\lambda5679.56)/I(\lambda5676.02)\simeq4.65$, so it is obtained $I(\lambda5679.56)=0.1177\pm0.0016$. Therefore, the line ratios would be $\lambda$5679.56/$\lambda$5666.63$\simeq$1.71 and    $\lambda5679.56$/$\lambda$4041.31$\simeq$1.12. While the former ratio indicates a range for $N_{e}$(N\,{\sc ii})=1000-3200\,cm$^{-3}$, the ratio falls outside the region that allows the determination of $T_{e}$ \citep[see Fig.\,8 of][]{fan11}. This inconsistency makes this range of $N_{e}$(N\,{\sc ii}) unsuitable to determine ionic abundances, even if used in combination with a $T_{e}$ determined with a different ion. Furthermore, as previously noted for the case of the CELs of [N\,{\sc ii}], the intensity of these lines may be affected by high-density zones (Zone 3) in the ionized gas, which could also affect the intensities of the N\,{\sc ii} ORLs, making these lines not suitable for determining physical conditions for both ORLs and CELs.

The relative intensity ratio $I(\lambda4649.13)/I(\lambda4591.63)$ from O\,{\sc ii} recombination lines is insensitive to electron density and has been used to determine the electron temperature \citep{fan13}. The O\,{\sc ii}$\lambda$4591 emission line is clearly detected in Vy\,2-2 in our spectra. However, the $\lambda$4649 emission line is a blend of O\,{\sc ii}$\lambda$4649.13, O\,{\sc ii}$\lambda$4650.84, and possibly, C\,{\sc iii} recombination lines. To ascertain the true contribution of $I(\lambda4649.13)$, we proceed as follows: Firstly, in our spectra, $I(\lambda4640)$ is a mix of O\,{\sc ii}$\lambda$4638.86, O\,{\sc ii}$\lambda$4641.81, and N\,{\sc ii}$\lambda$4640.64 intensities. By using the relative intensities from these ORLs reported in a recent high-resolution optical spectrum of Vy\,2-2 \citep{reb22}, we determine that $I(\lambda4638.86) + I(\lambda4641.81)$\,=\,0.564$\pm$0.015. Theoretically, $I(\lambda4641.81)/I(\lambda4638.86)$\,$\simeq$\,1.5. Thus, $I(\lambda4638.86)$\,=\,0.226$\pm$\,0.06 and $I(\lambda4641.81)$\,=\,0.338$\pm$0.09. Additionally, $I(\lambda4649.13)/I(\lambda4641.81)$\,$\simeq$\,1.1 and $I(\lambda4649.13)/I(\lambda4650.84)$\,$\simeq$\,1.5. Consequently, $I(\lambda4649.13)$\,=\,0.375$\pm$0.013 and $I(\lambda4650.84)$\,=\,0.225\,$\pm$\,0.013.
The total measured emission is $I(\lambda4649)$\,=\,0.985$\pm$0.019, including a $I(\lambda4650)$\,=\,0.385$\pm$0.026 component, likely originating from C\,\textsc{iii} recombination lines of stellar origin (Sec.\,\ref{central_star}). Lastly, from the O\,{\sc ii} recombination line intensity ratio $I(\lambda4649.13)/I(\lambda4591.63)$, we deduce $T_{e}$ (O\,{\sc ii})\,=\,6309$^{+769}_{-686}$ K. \cite{reb22} found $T_{e}$\,=\,8300$\pm$300\,K via a different method, consistent with our results.

The relative intensity ratio between $I(\lambda4075.86)/I(\lambda4070)$ of O\,{\sc ii} recombination lines is largely unaffected by electron temperature, making it useful for electron density determination \citep{fan13}. Specifically, $I(\lambda4070)$ is defined as $I(\lambda4069.89) + I(\lambda4069.62)$. In our measurements, this encompasses $I(\lambda4069.89) + I(\lambda4069.62) + I(\lambda4072.16)$. The theoretical intensity ratio $I(\lambda4069.89) + I(\lambda4069.62)/I(\lambda4072.16)$ is $\simeq$1.19. From our spectra, we find $I(\lambda4069)$=1.100$\pm$0.033, leading to $I(\lambda4070)$=0.592$\pm$0.018. Conversely, $I(\lambda4077) = I(\lambda4075.85) + I(\lambda4078.84)$ gives a value of 0.59$\pm$0.05, while theoretically, $I(\lambda4075.86)/I(\lambda4078.84)$$\simeq$7.09, allowing us to deduce $I(\lambda4075.86)$=0.51$\pm$0.05. Consequently, utilizing the O\,{\sc ii} recombination intensity ratio $I(\lambda4075.86)/I(\lambda4070)$, we derive an electron density $N_{e}$(O\,{\sc ii})\,=\,4467$^{+781}_{-664}$\,cm$^{-3}$. This differs substantially from the prior report by \cite{reb22}, which gave $N_{e}$(O\,{\sc ii})\,=\,19900$\pm$4300\,cm$^{-3}$ based on the $I(\lambda4649.13)/I(\lambda4661)$ ratio. This discrepancy most probably arises due to the fact that stellar C\,{\sc iii} emission lines are contaminating the nebular $I(\lambda4649.13)$ in Vy\,2-2, as analyzed in Sec.\,\ref{central_star}.

From the analysis of O\,{\sc ii} recombination lines, both the relative intensities between $I(\lambda4649.13)/I(\lambda4661.63)$ and $I(\lambda4649.13)/I(\lambda4089.29)$ can be use to determine physical conditions \citep{mcn13, fan13}. However, our CAFOS spectra have a limited spectral resolution and the $I(\lambda4661.63)$ is not detected or is blended with a much more intense $\lambda$4658 emission line. Therefore we cannot estimate $I(\lambda4661.63)$. Furthermore, the line ratio $I(\lambda4649.13)/I(\lambda4089.29)$ $\simeq$ 1.29 is below the range covered by the calculations presented in \cite{mcn13}.

We were not able to estimate an electron temperature from the Balmer jump. It lies at the very blue end of our spectra and even in the deepest spectrum the measurement is extremely inaccurate.

\section{Kinematical components in the ionized gas}

Table\,\ref{table:CARMENES} shows the kinematical components C1 to C4 detected in our CARMENES spectra (see Sec.\,\ref{sec:kinematics}) in selected emission lines, for which we report their radial velocity ($\upsilon_{r}$) with respect to the LSR, and their velocity width [$\Delta \upsilon$ (FWHM)].

\begin{table*}
	\caption{Radial velocity of the intensity peak [$\upsilon_{r}$ (LSR)] and velocity width [$\Delta \upsilon$ (FWHM)] of the kinematical components identified in the CARMENES spectra of Vy\,2-2 for selected emission lines. The line identification includes the usual air wavelength and we also provide the vacuum wavelength used for the calculations (see Sect.\,2.2 for details).} % title of Table
	\label{table:CARMENES}      % is used to refer this table in the text
	\centering                          % used for centering table
	\begin{tabular}{l c c c l c c c}        % centered columns (4 columns)
		\hline
		\rule{0pt}{2.5ex}                 % inserts double horizontal lines
		Line+$\lambda$(air)	& Kinematical	& $\upsilon_{r}$ (LSR) & $\Delta \upsilon$ (FWHM) & 	Line+$\lambda$(air)	& Kinematical	& $\upsilon_{r}$ (LSR) & $\Delta \upsilon$ (FWHM)\\
	$\lambda$(vacuum)($\AA$)	& component\tablefootmark{a} & (km\,s$^{-1}$)		&  (km\,s$^{-1}$) &	$\lambda$(vacuum)($\AA$) & component\tablefootmark{a} & (km\,s$^{-1}$)		&  (km\,s$^{-1}$) \\
		\hline                        % inserts single horizontal line
		\rule{0pt}{3ex}										
[N\,{\sc ii}]\,$\lambda$5755	&	C1	&	$-$75.8	&	18.4	&	[Fe\,{\sc ii}]\,$\lambda$7155	&	C1	&	$-$81.3	&	37.4 \\
$\lambda$5756.19	&	C2	&	$-$59.2	&	14.3	&	$\lambda$7157.14	&	C4	&	$-$17.2	&	48.7 \\
	&	C3	&	$-$43.5	&	17.0	&	O\,{\sc i}\,$\lambda$7254	&	C2	&	$-$56.9	&	9.3 \\
	&	C4	&	$-$26.9	&	40.0	&	$\lambda$7256.44	&	C3	&	$-$39.5	&	7.2 \\
He\,{\sc i}\,$\lambda$5874	&	C1	&	$-$80.4	&	29.0	&	He\,{\sc i}\,$\lambda$7281	&	C1	&	$-$72.6	&	32.9 \\
 $\lambda$5877.25	&	C2	&	$-$60.6	&	21.5	&	$\lambda$7283.36   &	C2	&	$-$58.6	&	20.7 \\
	&	C3	&	$-$41.7	&	22.1	&		&	C3	&	$-$40.1	&	13.6 \\
	&	C4	&	$-$22.3	&	28.4	&		&	C4	&	$-$29.0	&	21.9 \\

[O\,{\sc i}]\,$\lambda$6300\tablefootmark{b} &	C1	&	$-$81.7	&	30.5	&	[Ni\,{\sc ii}]\,$\lambda$7378	&	C1	&	$-$81.3	&	31.3 \\
$\lambda$6302.05	&	C2--C3	&	$-$50.3	&	29.0	&	$\lambda$7379.86	&	C4	&	$-$16.7	&	38.9 \\
	&	C4	&	$-$19.8	&	28.6	&	[Ar\,{\sc iii}]\,$\lambda$7751	&	C1	&	$-$76.8	&	25.4 \\
 
[S\,{\sc iii}]\,$\lambda$6312	&	C1	&	$-$73.6	&	28.3	&	$\lambda$7753.24	&	C2	&	$-$60.6	&	16.8 \\
$\lambda$6313.81	&	C2	&	$-$58.9	&	17.1	&		&	C3	&	$-$42.8	&	17.3 \\
	&	C3	&	$-$42.3	&	14.1	&		&	C4	&	$-$29.2	&	31.5 \\
	&	C4	&	$-$29.0	&	24.1	&	P\,5 $\lambda$8545	&	C1	&	$-$82.8	&	30.5 \\

[N\,{\sc ii}]\,$\lambda$6548	&	C1	&	$-$79.8	&	34.6	&	$\lambda$8547.31 &	C2	&	$-$59.5	&	29.0 \\
$\lambda$6549.86	&	C2	&	$-$60.2	&	20.0	&		&	C3	&	$-$40.8	&	28.6 \\
	&	C3	&	$-$40.9	&	16.8	&		&	C4	&	$-$15.0	&	27.9 \\
	&	C4	&	$-$21.7	&	33.4	&	[Fe\,{\sc ii}]\,$\lambda$8617	&	C1	&	$-$85.1	&	38.3 \\
H$\alpha$\,$\lambda$6563	&	C2	&	$-$59.2	&	46.6	&	$\lambda$8619.32	&	C4	&	$-$22.8	&	57.0 \\
$\lambda$6564.63	&	C3	&	$-$46.4	&	39.3	&	He\,{\sc i}\,$\lambda$11970	&	C1	&	$-$86.2	&	14.8 \\

[N\,{\sc ii}]\,$\lambda$6583	&	C1	&	$-$80.2	&	31.5	&	$\lambda$11972.32 &	C2	&	$-$61.2	&	19.8 \\
$\lambda$6585.27	&	C2	&	$-$61.1	&	19.3	&		&	C3	&	$-$38.6	&	17.9 \\
	&	C3	&	$-$42.4	&	16.8	&		&	C4	&	$-$26.1	&	31.8 \\
	&	C4	&	$-$23.8	&	32.0	&	[Fe\,{\sc ii}]\,$\lambda$12567	&	C1	&	$-$85.6	&	25.1 \\
He\,{\sc i}\,$\lambda$6678	&	C1	&	$-$82.2	&	26.0	&	$\lambda$12570.20	&	C2	&	$-$64.2	&	27.0 \\
$\lambda$6680.00	&	C2	&	$-$61.6	&	21.2	&		&	C3	&	$-$33.2	&	26.2 \\
	&	C3	&	$-$42.7	&	21.8	&		&	C4	&	$-$9.3	&	18.8 \\
	&	C4	&	$-$23.4	&	29.1	&	P$\beta$ $\lambda$12818	&	C1	&	$-$81.6	&	27.4 \\
He\,{\sc i}\,$\lambda$7065	&	C1	&	$-$78.9	&	24.0	&	$\lambda$12821.58	&	C2	&	$-$59.5	&	28.9 \\
$\lambda$7067.14	&	C2	&	$-$61.9	&	17.8	&		&	C3	&	$-$42.6	&	29.1 \\
	&	C3	&	$-$43.2	&	20.6	&		&	C4	&	$-$20.5	&	27.8 \\
	&	C4	&	$-$25.8	&	29.5	&	[Fe\,{\sc ii}]\,$\lambda$16435	&	C1	&	$-$84.8	&	25.9 \\

[Ar\,{\sc iii}]\,$\lambda$7135	&	C1	&	$-$75.9	&	26.1	& $\lambda$16440.01	&	C2	&	$-$57.4	&	25.8 \\
$\lambda$7137.76	&	C2	&	$-$60.3	&	16.6	&		&	C3	&	$-$33.7	&	25.1 \\
	&	C3	&	$-$43.1	&	15.8	&		&	C4	&	$-$11.83	&	25.6 \\
	&	C4	&	$-$28.8	&	23.9	&		\\					
		\hline                                   %inserts single line
	\end{tabular}
\tablefoot{
  \tablefootmark{a}{C1 and C4 represent blueshifted and redshifted emission, respectively, from the bipolar lobes; C2 and C3 represent blueshifted and redshifted emission, respectively, from the central region (see the text and Fig.\,4 for details).}
  \tablefootmark{b}{C2 and C3 are not separated in radial velocity in this emission line (see the text).  }
  }
\end{table*}

\end{appendix}

\end{document}